\documentclass[prb,twocolumn,showpacs,superscriptaddress,nofootinbib]{revtex4}

\usepackage{epsfig}
\usepackage{amsmath,amssymb}
\usepackage{graphicx}
\usepackage[noabbrev]{cleveref}
\usepackage{caption}
\usepackage{adjustbox}
\usepackage{floatrow}
\usepackage{subfig}
\usepackage{footnote}
\usepackage{grffile}
\usepackage{color}
\usepackage{dsfont}
\usepackage{mathtools}
\usepackage{verbatim}
\usepackage{bbold}
\usepackage{hhline}

\allowdisplaybreaks

\newcommand{\dbunderline}[1]{\underline{\underline{#1}}}

\begin{document}

\title{Efficient Bethe-Salpeter equations' treatment in dynamical mean-field theory}

\newcommand{\TUVienna}{\affiliation{Institute for Solid State Physics, Vienna University of Technology, 1040 Vienna, Austria}}
\newcommand{\Uniwien}{\affiliation{Physics of Nanostructured Materials, Faculty of Physics, University of Vienna, 1090 Vienna, Austria}}
\newcommand{\UniTueb}{\affiliation{Institut f\"ur Theoretische Physik and Center for Quantum Science, Universit\"at T\"ubingen, Auf der Morgenstelle 14, 72076 T\"ubingen, Germany}}
\newcommand{\RussianQC}{\affiliation{Russian Quantum Center, 143025 Skolkovo, Russia}}
\newcommand{\Saclay}{\affiliation{Institut de Physique Th\'{e}orique (IPhT), CEA, CNRS, 91191 Gif-sur-Yvette, France}}

\newcommand{\MoscowUni}{\affiliation{Department of Physics, M.V. Lomonosov Moscow State University, 119991 Moscow, Russia}}

\author{Agnese Tagliavini} 	\TUVienna \UniTueb

\author{Stefan Hummel} 		\Uniwien

\author{Nils Wentzell} 	    \TUVienna  \Saclay

\author{Sabine Andergassen} 	\UniTueb

\author{Alessandro Toschi}  	\TUVienna

\author{Georg Rohringer} 	\RussianQC

\begin{abstract}
 We present here two alternative schemes designed to correct the high-frequency truncation errors in the numerical treatment of the Bethe-Salpeter equations. The schemes are applicable to all Bethe-Salpeter calculations with a local two-particle irreducible vertex, which is relevant, e.g., for the dynamical mean-field theory (DMFT) and its diagrammatic extensions. In particular, within  a purely diagrammatic framework, we could extend existing algorithms for treating the static case in the particle-hole sector to more general procedures applicable to {\sl all bosonic frequencies} and {\sl all} channels. After illustrating the derivation and the theoretical interrelation of the two proposed schemes, these have been applied to the Bethe-Salpeter equations for the auxiliary Anderson impurity models of selected  DMFT calculations, where results can be compared against a numerically ``exact" solution. The successful performance of the proposed schemes suggests that their implementation can significantly improve the accuracy of DMFT calculations at the two-particle level, in particular for more realistic multi-orbital calculations where the large number of degrees of freedom substantially restricts the actual frequency range for numerical calculations, as well as -on a broader perspective- of the diagrammatic extensions of DMFT.


\end{abstract}

\maketitle

\section{Introduction}
\label{sec:introduction}

Scattering experiments have always been of utmost importance for gaining new insights into the laws of physics. Among the most famous findings which have been achieved by this technique are the unveiling of the structure of the atom \cite{Rutherford1911} or the discovery of the Higgs boson \cite{Aad2012} which confirmed one of the major predictions of the standard model of particle theory. On the theoretical side, Feynman diagrammatic perturbation theory turned out to be a powerful tool to calculate the corresponding two-particle scattering amplitudes. This technique has been particularly successful in theories such as quantum electrodynamics where a small expansion parameter (like the fine structure constant $\alpha$) allows to truncate the perturbation series at low order.

The situation is, however, very different for highly correlated (lattice) electrons where the strong interaction between the particles prevents any finite-order perturbative treatment. In fact, a rich spectrum of correlation-driven phenomena, like the celebrated Mott metal-to-insulator transition\cite{Mott1968,McWhan1969} (MIT) or the high-temperature superconductivity in the cuprates\cite{Bednorz1986,Lee2006}, are inaccessible by finite diagrammatic expansion. In this respect, significant progress has been achieved by the dynamical mean field theory\cite{Metzner1989,Georges1992,Georges1996} (DMFT) which can describe all purely local correlations in the system. DMFT has several successes, like the understanding of realistic correlated materials\cite{Kotliar2006,Held2007}. However, it still exhibits a number of limitations which prevent a complete description of (realistic) correlated electron systems. Two of them are of particular importance:

(i) Hitherto, DMFT calculations have been often restricted to the one-particle level, i.e., to the calculation of the self-energy and the spectral function of the system. While the latter provide crucial information about the one-particle excitations and can be compared to (angular resolved) photoemission spectroscopy (ARPES), the insights that can be gained from two-particle correlation functions are certainly of equal (or even higher) significance. Among the latter are the optical conductivity and the magnetic, charge, or particle-particle susceptibility which describe the linear response of the system to an external (electric, magnetic or pairing) source field. In the case of strong correlations, an accurate evaluation of these response functions requires the inclusion of vertex corrections\cite{Toschi2012,Pavarini2014}. These have been in most cases neglected with the exception of few recent calculations \cite{BoehnkePhd2015, Steiner2016, Hoshino2014, Hoshino2015, Hoshino2016}. Moreover, an improved treatment of two-particle vertex functions will be also highly beneficial for the so-called "fluctuation diagnostics" method\cite{Gunnarsson2015}, which has been applied to the self-energy or spectral function to disentangle features originated from different collective fluctuations in the system.

(ii) An intrinsic limitation of DMFT is the locality of the self-energy which, hence, includes only local correlations effects. However, neglecting the impact of non-local collective modes on the spectral function yields often poor results at low temperature where long-range order parameter fluctuations dominate the physics. In order to overcome these difficulties, so-called cluster, e.g., cellular DMFT\cite{Kotliar2001} and dynamical cluster approximation\cite{Hettler1998,Hettler2000,Maier2005} (DCA) , as well as diagrammatic extensions of DMFT\cite{Rohringer2018} have been developed. The most important among the latter are the diagrammatic vertex approximation\cite{Toschi2007,Toschi2011,Rohringer2016} (D$\Gamma$A), the dual fermion (DF) theory\cite{Rubtsov2008}, the one-particle irreducible approach\cite{Rohringer2013} (1PI), the dual boson (DB) theory\cite{Rubtsov2012}, DMF$^{2}$RG\cite{Taranto2014} as well as the  TRILEX\cite{Ayral2015,Ayral2016a} and the QUADRILEX\cite{Ayral2016} method.

Both (i) the calculation of vertex corrections within DMFT and (ii) the inclusion of non-local correlation effects beyond DMFT in the self-energy require the calculation of (non-local) ladder diagrams. This is achieved by DMFT building blocks, i.e., the non-local DMFT Green's function and the (local) two-particle irreducible vertex (in a given channel), which makes an accurate determination of the latter indispensable. In practice, this local irreducible vertex is obtained from inverting the so-called (local) Bethe-Salpeter (BS) equations which are integral equations with respect to (w.r.t.) the fermionic Matsubara frequency arguments of the local one- and two-particle DMFT correlation functions. While, in principle, the ``internal" fermionic frequency sums run on an infinite range, in practice, one has to restrict oneself to a finite frequency interval. This transforms the inversion of the BS equations into the inversion of a finite matrix in the fermionic frequency space, whose size is, due to the numerical computational cost and possibly memory constraints, rather small. This limitation is particularly severe when the parquet formalism\cite{Bickers2004,Janis2001,Rohringer2012} is used. Obviously, such a cut-off procedure in the frequency space produces an error which might drastically limit the accuracy of two-particle response functions or results of diagrammatic  extensions of DMFT based on these irreducible vertices.

In this paper, we present new schemes to correct the errors which arise from the restriction of the BS equations to a finite frequency grid. The proposed procedures are based on the observation that the complexity of the frequency dependence of the DMFT two-particle vertex functions is drastically reduced for large values of these frequencies \cite{Rohringer2012,Li2016,Wentzell2016a}. In fact, in the asymptotic high-frequency regime the DMFT scattering amplitude can be represented by related response functions and fermion-boson vertices which depend on one and two (instead of three) frequencies, respectively. Hence, they can be evaluated numerically on a much larger frequency grid, and we can approximate the full two-particle correlation functions of DMFT by these so-called asymptotic functions outside the (possibly rather small) frequency regime where the full three-frequency objects have to be calculated. A similar procedure has been already proposed in Ref.~[\onlinecite{Kunes2011}] where, however, the asymptotic functions have been obtained by means of functional derivatives of the normal self-energy, restricting the method to BS equations in the particle-hole channels at bosonic transfer frequency $\omega\!=\!0$. Here, instead, we adopt a diagrammatic analysis of the two-particle vertices of DMFT, put forward in Refs.~[\onlinecite{Rohringer2012,Li2016,Wentzell2016a}], which provides the high-frequency behavior of the vertex for {\sl all} channels and also for $\omega\!\ne\!0$. Besides extending the method presented in Ref.~[\onlinecite{Kunes2011}], this also allows us to develop a new approach based on the asymptotics of the full two-particle correlations functions.

The paper is organized in the following way. In Sec.~\ref{sec:theory} we give the explicit expressions of all local DMFT two-particle correlations functions used throughout the paper and discuss their high-frequency asymptotic behavior. This is then used in Sec.~\ref{sec:numerical_impl} for the two newly proposed methods for accurately solving the BS equations. In Sec.~\ref{sec:num_results} numerical results are presented for both approaches and Sec.~\ref{sec:conclusions} is devoted to conclusions and an outlook.

\section{Theory and formalism}
\label{sec:theory}

In this section, we will give all necessary definitions and recall the asymptotic high-frequency behavior of local two-particle correlation functions which are relevant for the main goal of this paper, i.e., a consistent and numerically stable calculation of the irreducible vertex functions of DMFT. More specifically, in Sec.~\ref{subsec:definitions}, we introduce the model, for which the applicability of our new approach has been tested, and the general two-particle formalism which is required for the development of our new methods. In Sec.~\ref{subsec:asympt}, we revisit\cite{Rohringer2012,Rohringer2014,Wentzell2016a} the diagrammatic techniques for analyzing the frequency structure of the different two-particle vertex functions and present their high-frequency behavior in terms of collective modes (described by physical susceptibilities) and their interaction with the particles of the system (fermion-boson vertex).

\subsection{Definitions and notation}
\label{subsec:definitions}

We consider the single-band Hubbard model for a generic lattice in $d$ dimensions,
\begin{equation}
 \label{eq:defhamilt}
 \hat{\mathcal{H}}=-t\sum_{\langle ij \rangle , \sigma}\hat{c}^{\dagger}_{i\sigma}\hat{c}_{j\sigma}+U\sum_i \hat{n}_{i\uparrow}\hat{n}_{i\downarrow} - \mu \sum_{i,\sigma}\hat{n}_{i\sigma} ,
\end{equation}
where $\hat{c}^{(\dagger)}_{i\sigma}$ annihilates (creates) an electron with spin $\sigma$ at the lattice site $\mathbf{R}_i$ ($\hat{n}_{i\sigma}=\hat{c}^{\dagger}_{i\sigma}\hat{c}_{i\sigma}$), $t$ is the hopping amplitude for electrons between neighboring sites, $\mu$ the chemical potential and $U$ the on-site Coulomb interaction. Here, we adopt DMFT to treat this model and, hence, we consider the purely local one- and two-particle correlation and (reducible as well as irreducible) vertex functions of the Anderson impurity model (AIM) related to the DMFT solution of the Hamiltonian given in Eq.~(\ref{eq:defhamilt}). The basic two-particle correlation function, from which the irreducible vertices can be derived, is the generalized susceptibility defined by
\begin{equation}
\label{equ:defchi}
\begin{split}
 \chi^{\nu\nu'\omega}_{ph,\sigma\sigma'}&=\int\limits_{0}^{\beta}{d\tau_{1}d\tau_{2}d\tau_{3} \, e^{-i\nu\tau_{1}}e^{i(\nu+\omega)\tau_{2}}e^{-i(\nu'+\omega)\tau_{3}}} \\&\times \big[\big<T_{\tau}c_{\sigma}^{\dagger}(\tau_{1})c_{\sigma}(\tau_{2})c_{\sigma'}^{\dagger}(\tau_{3})c_{\sigma'}(0)\big> \\&- \big<T_{\tau}c_{\sigma}^{\dagger}(\tau_{1})c_{\sigma}(\tau_{2})\big>\big<T_{\tau}c_{\sigma'}^{\dagger}(\tau_{3})c_{\sigma'}(0)\big>\big],
\end{split}
\end{equation}
where $\nu(\nu')\!=\!\frac{\pi}{\beta}(2n(n')\!+\!1)$, $n,n'\!\in\mathds{Z}$, is a fermionic and $\omega\!=\!\frac{\pi}{\beta}2m$, $m\!\in\!\mathds{Z}$, a bosonic Matsubara frequency. $T_{\tau}$ is the time-ordering operator and $\langle\ldots\rangle\!=Z^{-1}\mbox{Tr}(e^{-\beta \hat{\mathcal{H}}}\ldots)$ denotes the thermal expectation value, with $Z\!=\!\mbox{Tr}(e^{-\beta \hat{\mathcal{H}}})$. $\beta\!=\!1/T$ is the inverse temperature of the system. The assignment of the frequencies $\nu$, $\nu+\omega$ and $\nu'+\omega$ to the imaginary times $\tau_1$, $\tau_2$ and $\tau_3$, respectively, corresponds to the so-called particle-hole ($ph$) notation\cite{Rohringer2012}. Analogously, one can express the generalized susceptibility in the transverse particle-hole ($\overline{ph}$) or in the particle-particle ($pp$) notation which can be obtained from the $ph$ one by a mere frequency shift, i.e., $\chi^{\nu\nu'\omega}_{\overline{ph},\sigma\sigma'}\!\equiv\!\chi^{\nu(\nu+\omega)(\nu'-\nu')}_{{ph},\sigma\sigma'}$ and $\chi^{\nu\nu'\omega}_{{pp},\sigma\sigma'}\!\equiv\!\chi^{\nu\nu'(\omega-\nu-\nu')}_{{ph},\sigma\sigma'}$, respectively. The different physical interpretations of these notations as particle-hole and particle-particle scattering amplitude are discussed in detail in Refs.~[\onlinecite{Rohringer2012}] and [\onlinecite{Wentzell2016a}] (see, in particular, Figs.~1 and 2 in the former).

In the SU(2) symmetric situation considered here, it is convenient to decompose the two-particle correlation functions into their spin singlet- and spin triplet-components, both for the $ph$ and the $pp$ representation. This corresponds to the definitions of the generalized susceptibilities in the density ($d$), magnetic ($m$), particle-particle singlet ($s$) and particle-particle triplet ($t$) channels\footnote{Note that the definitions for the singlet($s$) and triplet($t$) susceptibilities slightly differ from the corresponding ones given in Ref.~[\onlinecite{Rohringer2012}] (Eqs.~B19 therein) in order to obtain a unified form for the BS equation in all channels.}:
\begin{subequations}
\label{equ:defsuscchannel}
\begin{align}
\label{equ:defsuscchanneld}
 &\chi_d^{\nu\nu'\omega}=\chi_{ph,\uparrow\uparrow}^{\nu\nu'\omega}+\chi_{ph,\uparrow\downarrow}^{\nu\nu'\omega},\\
\label{equ:defsuscchannelm}
&\chi_m^{\nu\nu'\omega}=\chi_{ph,\uparrow\uparrow}^{\nu\nu'\omega}-\chi_{ph,\uparrow\downarrow}^{\nu\nu'\omega},\\
\label{equ:defsuscchannels}
&\chi_s^{\nu\nu'\omega}=\frac{1}{4}(-\chi_{pp,\uparrow\uparrow}^{\nu\nu'\omega}+2\chi_{pp,\uparrow\downarrow}^{\nu\nu'\omega}-2\chi_{0,pp}^{\nu\nu'\omega}),\\
\label{equ:defsuscchannelt}
&\chi_t^{\nu\nu'\omega}=\frac{1}{4}(\chi_{pp,\uparrow\uparrow}^{\nu\nu'\omega}+2\chi_{0,pp}^{\nu\nu'\omega}).
\end{align}
\end{subequations}
The bare susceptibilities in the particle-hole and particle-particle notation are given by
\begin{subequations}
\label{equ:defsuscbare}
\begin{align}
\label{equ:defsuscbareph}
&\chi_{0,d/m}^{\nu\nu'\omega}=\chi_{0,ph}^{\nu\nu'\omega}=-\beta G(\nu)G(\nu+\omega)\delta_{\nu\nu'},\\
\label{equ:defsuscbarepp}
&\chi_{0,s/t}^{\nu\nu'\omega}=\chi_{0,pp}^{\nu\nu'\omega}=-\frac{\beta}{2} G(\nu)G(\omega-\nu)\delta_{\nu\nu'},
\end{align}
\end{subequations}
where $G(\nu)$ is the local single-particle DMFT Green's function. From the generalized susceptibilities in Eqs.~(\ref{equ:defsuscchannel}) the corresponding physical ones can be obtained by summing the former over the fermionic Matsubara frequencies $\nu$ and $\nu'$. They explicitly read [see Fig.~\ref{fig:diagrams}(a)]
\begin{align}
\label{equ:susphysicdm}
 &\chi_{d/m}^{\omega}=\frac{1}{\beta^2}\sum_{\nu\nu'}\chi_{d/m}^{\nu\nu'\omega},\nonumber\\
 &\chi_{pp,\uparrow\downarrow}^{\omega}=\frac{2}{\beta^2}\sum_{\nu\nu'} \chi_s^{\nu\nu'\omega}=\frac{1}{\beta^2}\sum_{\nu\nu'}\left(\chi_{pp,\uparrow\downarrow}^{\nu\nu'\omega}-2\chi_{0,pp}^{\nu \nu' \omega}\right),
\end{align}
and describe the physical response to a (local) chemical potential, a magnetic and a singlet ($\uparrow\downarrow$) pairing field, respectively. The last equality in Eq.~(\ref{equ:susphysicdm}) follows from $\sum_{\nu\nu'}(\chi_t^{\nu\nu'\omega}-\chi_{0,pp}^{\nu\nu'\omega})\!=\!0$, which is a consequence of the Pauli principle and reflects the fact, that in a purely local (single-orbital) model no triplet superconductivity is possible.

From the definition of the generalized susceptibilities a number of different {\sl vertex} functions can be derived. By removing all unconnected parts and amputating the outer legs from $\chi_r^{\nu\nu'\omega}$ the so-called {\sl full} two-particle vertex is obtained as
\begin{equation}
\label{equ:defF}
 F_r^{\nu\nu'\omega}=-\frac{\chi_r^{\nu\nu'\omega}\mp\chi_{0,r}^{\nu\nu'\omega}}{\frac{1}{\beta^2}\sum_{\nu_1\nu_2}\chi_{0,r}^{\nu\nu_1\omega}\chi_{0,r}^{\nu_2\nu'\omega}},
\end{equation}
where the minus sign has to be used for $r\!=\!d,m,t$ and the plus sign for $r\!=\!s$. Physically, $F_r^{\nu\nu'\omega}$ corresponds the full two-particle scattering amplitude between (quasi)particles \cite{Abrikosov1975}, which is represented by the set of {\sl all} connected (amputated) two-particle Feynman diagrams.

Diagrammatically, by ``gluing'' together the outgoing (or the incoming) outer legs of a generalized susceptibility (i.e., summing over the respective fermionic frequency) and amputating the two remaining outer legs, one obtains the fermion-boson vertices\footnote{Note that the fermion-boson vertices in Eqs.~(\ref{equ:fermbosvert}) differ from the corresponding definitions in the dual boson\cite{Rubtsov2012} and TRILEX\cite{Ayral2015,Ayral2016a} theories by a factor $1\pm U\chi_r^{\omega}$. Physically, this means that the $\lambda_r^{\nu\omega}$'s as given in Eqs.~(\ref{equ:fermbosvert}) still contain contributions from collective modes\cite{Rohringer2016}.} [see also Fig.~\ref{fig:diagrams}(b)]
\begin{align}
\label{equ:fermbosvert}
 \lambda_{d/m}^{\nu\omega}&=\mp\frac{1}{\beta}\frac{\sum_{\nu'}\chi_{d/m}^{\nu\nu'\omega}}{G(\nu)G(\nu+\omega)}\mp 1,\nonumber\\
 \lambda_{pp,\uparrow\downarrow}^{\nu\omega}&=\frac{2}{\beta}\frac{\sum_{\nu'}\chi_{s}^{\nu\nu'\omega}}{G(\nu)G(\omega-\nu)}-1=\frac{1}{\beta}\frac{\sum_{\nu'}\chi_{pp,\uparrow\downarrow}^{\nu\nu'\omega}}{G(\nu)G(\omega-\nu)},
\end{align}
and $\lambda_r^{\nu'\omega}$ is obtained by exchanging $\nu$ and $\nu'$. The fermion-boson vertices $\lambda_r^{\nu\omega}$ are related to the interaction between a fermion with energy $\nu$ and a collective charge, spin or singlet ($\uparrow\downarrow$) particle-particle excitations with frequency $\omega$, respectively. The last equality in Eq.~(\ref{equ:fermbosvert}) as well as the vanishing of a triplet particle-particle fermion-boson vertex, i.e., $\lambda_{t=\uparrow\uparrow}^{\nu\omega}\!=\!0$, again follow from the Pauli principle.

Let us finally turn our attention to the central objects of this paper, i.e., the vertices $\Gamma_r^{\nu\nu'\omega}$ which are two-particle irreducible in channel $r$. A Feynman diagram for the two-particle vertex is called irreducible in channel $r=ph,\overline{ph},pp$, if it {\sl cannot} be split into two separated diagrams by cutting two internal fermionic lines in such a way that one of the two contains the outer frequencies $(\nu,\pm\nu+\omega)$ and the other $(\nu',\pm\nu'+\omega)$ w.r.t. the corresponding natural frequency convention [see discussion below Eq.~(\ref{equ:defchi})]. The sum of all diagrams of a certain type then yields the irreducible vertices $\Gamma_{ph,\sigma\sigma'}^{\nu\nu'\omega}$, $\Gamma_{\overline{ph},\sigma\sigma'}^{\nu\nu'\omega}$, and $\Gamma_{pp,\sigma\sigma'}^{\nu\nu'\omega}$ (which are always assumed to be represented in their corresponding frequency notation). Due to the crossing and SU(2) symmetry\cite{Bickers2004,Rohringer2012}, $\Gamma_{\overline{ph},\sigma\sigma'}^{\nu\nu'\omega}$ can be expressed in terms of $\Gamma_{ph,\sigma\sigma'}^{\nu\nu'\omega}$ and, hence, we can restrict ourselves to $\Gamma_{ph,\sigma\sigma'}^{\nu\nu'\omega}$ and $\Gamma_{pp,\sigma\sigma'}^{\nu\nu'\omega}$ in the following. Moreover, in the SU(2) symmetric case, where $\uparrow\uparrow\!=\!\downarrow\downarrow$ and $\uparrow\downarrow\!=\!\downarrow\uparrow$, it is convenient to introduce spin-singlet and spin-triplet components for the irreducible vertices\cite{Rohringer2012} [analogously as for the generalized susceptibilities in Eq.~(\ref{equ:defsuscchannel})]
\begin{subequations}
\label{equ:defgammaspindiag}
\begin{align}
&\Gamma_{d}^{\nu\nu'\omega}=\Gamma_{ph,\uparrow\uparrow}+\Gamma_{ph,\uparrow\downarrow},\label{defgammaspindiagd}\\
&\Gamma_{m}^{\nu\nu'\omega}=\Gamma_{ph,\uparrow\uparrow}-\Gamma_{ph,\uparrow\downarrow},\label{defgammaspindiagm}\\
&\Gamma_{s}^{\nu\nu'\omega}=-\Gamma_{pp,\uparrow\uparrow}+2\Gamma_{pp,\uparrow\downarrow},\label{defgammaspindiags}\\
&\Gamma_{t}^{\nu\nu'\omega}=\Gamma_{pp,\uparrow\uparrow},\label{defgammaspindiagt}
\end{align}
\end{subequations}
Note that here (in contrast to the corresponding definitions for $\chi_r^{\nu\nu'\omega}$), the index $r$ refers to {\sl both} the channel in which the vertex is irreducible ($ph$ for $r\!=\!d,m$ or $pp$ for $r\!=\!s,t$) as well as the spin combination and the frequency notation in which the vertex is represented ($\Gamma_{d,m}^{\nu\nu'\omega}$ in the $ph$ and $\Gamma_{s,t}^{\nu\nu'\omega}$ in the $pp$ frequency notation).

The irreducible vertex functions can be now obtained from the generalized susceptibilities via the BS equation
\begin{equation}
\label{equ:defBS}
\pm\chi_r^{\nu\nu'\omega}=\chi_{0,r}^{\nu\nu'\omega}-\frac{1}{\beta^2}\sum_{\nu_1\nu_2}\chi_{0,r}^{\nu\nu_1\omega}\Gamma_r^{\nu_1\nu_2\omega}\chi_r^{\nu_2\nu'\omega},
\end{equation}
with the plus for $r\!=\!d,m,t$ and the minus for $r\!=\!s$. By solving Eq.~(\ref{equ:defBS}) for the irreducible vertex $\Gamma_r^{\nu\nu'\omega}$ we obtain
\begin{equation}
\label{equ:calcGamma}
 \dbunderline{\Gamma_r^\omega}=\beta^2[(\dbunderline{\chi_r^\omega})^{-1}\mp(\dbunderline{\chi_{0,r}^\omega})^{-1}],
\end{equation}
with the minus for $r\!=\!d,m,t$ and the plus sign for $r\!=\!s$. $\dbunderline{X_r^\omega}\!\equiv\!X_r^{\nu\nu'\omega}$ indicates a (infinite) matrix in the fermionic frequencies $\nu$ and $\nu'$ (for a given value of the bosonic frequency $\omega$) and $(\dbunderline{X_r^\omega})^{-1}$ is its inverse (w.r.t. $\nu$ and $\nu'$).

\subsection{Asymptotics of the vertex functions}
\label{subsec:asympt}

In this section we review the behavior\cite{Rohringer2012,Rohringer2014,Wentzell2016a} of the irreducible vertex $\Gamma_r^{\nu\nu'\omega}$ and the generalized susceptibility $\chi_r^{\nu\nu'\omega}$ for large values of $\nu$ and $\nu'$ (for a fixed value of the bosonic Matsubara frequency $\omega$). As it has been discussed extensively in  Refs.~[\onlinecite{Rohringer2012,Hummel2014,Wentzell2016a}], this high-frequency asymptotics can be expressed in terms of the physical susceptibilities $\chi_r^{\omega}$ [Eq.~(\ref{equ:susphysicdm})] and the fermion-boson vertices $\lambda_r^{\nu\omega}$ [Eq.~(\ref{equ:fermbosvert})]. Here we present just the final expressions for $\Gamma_{r,\text{asym}}^{\nu\nu'\omega}$ and $\chi_{r,\text{asym}}^{\nu\nu'\omega}$ (or $F_{r,\text{asymp}}^{\nu\nu'\omega}$) and refer the reader to Appendix~\ref{app:derivasympt} for their explicit derivations.

The vertex $\Gamma_r^{\nu\nu'\omega}$ contains all diagrams which are not reducible in channel $r$. Its asymptotic high-frequency behavior for $\nu,\nu'\!\rightarrow\!\infty$ and a fixed value of $\omega$ is, hence, determined by the fully irreducible vertex $\Lambda_r^{\nu\nu'\omega}$ and the reducible vertices $\Phi_{r'}^{\nu\nu'\omega}$ with $r'\!\ne\!r$ [see parquet Eqs.~(\ref{eq:parquet_eqs}) in Appendix~\ref{app:derivasympt}]. The former contributes just via the bare interaction, i.e., by a term $\propto U$ (where the prefactor depends on the channel $r$), while the high-frequency behavior of the latter can be expressed in terms of the physical susceptibilities $\chi_r^{\omega}$ [Eq.~(\ref{equ:susphysicdm})]. Following the discussion in Appendix~\ref{app:derivasympt}, we explicitly obtain for $\Gamma_{r,\text{asym}}^{\nu\nu'\omega}$,
\begin{widetext}
\begin{subequations}
\label{equ:Gammahighfreq}
\begin{align}
 \label{equ:Gammadhighfreq}
 &\Gamma_{d,\text{asym}}^{\nu\nu'\omega}=U+\frac{U^2}{2}\chi_d^{\nu'-\nu}+\frac{3U^2}{2}\chi_m^{\nu'-\nu}-U^2\chi_{pp,\uparrow\downarrow}^{\nu+\nu'+\omega},\\
 \label{equ:Gammamhighfreq}
 &\Gamma_{m,\text{asym}}^{\nu\nu'\omega}=-U+\frac{U^2}{2}\chi_{d}^{\nu'-\nu}-\frac{U^2}{2}\chi_{m}^{\nu'-\nu}+U^2\chi_{pp,\uparrow \downarrow}^{\nu+\nu'+\omega},\\
 \label{equ:Gammashighfreq}
 &\Gamma_{s,\text{asym}}^{\nu\nu'\omega}=2U-\frac{U^2}{2}\chi_{d}^{\nu'-\nu}+\frac{3U^2}{2}\chi_{m}^{\nu'-\nu}-\frac{U^2}{2}\chi_{d}^{\omega-\nu-\nu'}+\frac{3U^2}{2}\chi_{m}^{\omega-\nu-\nu'},\\
 \label{equ:Gammathighfreq}
 &\Gamma_{t,\text{asym}}^{\nu\nu'\omega}=\frac{U^2}{2}\chi_{d}^{\nu'-\nu}+\frac{U^2}{2}\chi_{m}^{\nu'-\nu}-\frac{U^2}{2}\chi_{d}^{\omega-\nu -\nu'}-\frac{U^2}{2} \chi_{m}^{\omega-\nu-\nu'}.
\end{align}
\end{subequations}
\end{widetext}
We note that this asymptotic expansion corresponds to the one presented in Ref.~[\onlinecite{Wentzell2016a}], where the susceptibilities $\chi_r^{\omega}$ (multiplied by $U^2$) have been referred to as Kernel-one functions $\mathcal{K}_{1,r}^\omega$.

In order to obtain the high-frequency asymptotics of the full vertex we use that $F_r^{\nu\nu'\omega}=\Gamma_r^{\nu\nu'\omega}\!+\Phi_r^{\nu\nu'\omega}$. Hence, we only have to add the high-frequency contributions of $\Phi_r^{\nu\nu'\omega}$ to the corresponding ones of $\Gamma_r^{\nu\nu'\omega}$ in Eqs.~(\ref{equ:Gammahighfreq}). As discussed in detail in Appendix~\ref{app:derivasympt}, the former correspond to the fermion-boson vertex $\lambda_r^{\nu\omega}$ [see Eqs.~(\ref{equ:fermbosvert}) and (\ref{eq:phi_asy})], which leads to the following expressions for the high-frequency behavior of $F_{r}^{\nu\nu'\omega}$ for $\nu,\nu'\!\rightarrow\!\infty$ (with $\omega$ fixed):
\begin{widetext}
\begin{subequations}
\label{eq:f_eqs_asy}
\begin{align}
 \label{eq:fd_eqs_asy}
 &F_{d,\text{asym}}^{\nu\nu'\omega}=U+\frac{U^2}{2}\chi_d^{\nu'-\nu}+\frac{3U^2}{2}\chi_m^{\nu'-\nu}-U^2\chi_{pp,\uparrow\downarrow}^{\nu+\nu'+\omega}+U\lambda_d^{\nu\omega}+U\lambda_d^{\nu'\omega}+U^2\chi_d^{\omega},\\
 \label{eq:fm_eqs_asy}
 &F_{m,\text{asym}}^{\nu\nu'\omega}=-U+\frac{U^2}{2}\chi_{d}^{\nu'-\nu}-\frac{U^2}{2}\chi_{m}^{\nu'-\nu}+U^2\chi_{pp,\uparrow\downarrow}^{\nu+\nu'+\omega}+U\lambda_{m}^{\nu\omega}+U\lambda_{m}^{\nu'\omega}+U^2\chi_{m}^{\omega},\\
 \label{eq:fs_eqs_asy}
 &F_{s,\text{asym}}^{\nu\nu'\omega}=2U-\frac{U^2}{2}\chi_{d}^{\nu'-\nu}+\frac{3U^2}{2}\chi_{m}^{\nu'-\nu}-\frac{U^2}{2}\chi_{d}^{\omega-\nu-\nu'}+\frac{3U^2}{2}\chi_{m}^{\omega-\nu-\nu'}+2U\lambda_{pp,\uparrow\downarrow}^{\nu\omega}+2U\lambda_{pp,\uparrow\downarrow}^{\nu' \omega}+2U^2\chi_{pp,\uparrow\downarrow}^{\omega},\\
 \label{eq:ft_eqs_asy}
 &F_{t,\text{asym}}^{\omega}=\frac{U^2}{2}\chi_{d}^{\nu'-\nu}+\frac{U^2}{2}\chi_{m}^{\nu'-\nu}-\frac{U^2}{2}\chi_{d}^{\omega-\nu-\nu'}-\frac{U^2}{2}\chi_{m}^{\omega-\nu-\nu'}.
\end{align}
\end{subequations}
\end{widetext}
We note that $\lambda_r^{\nu\omega}$ is related to the so-called Kernel-two functions of Ref.~[\onlinecite{Wentzell2016a}] by $\mathcal{K}_{2,r}^{\nu\omega}\!\sim\!U\lambda_r^{\nu'\omega}\!+\!U^2\chi_r^{\omega}$.

From $F_{r,\text{asym}}^{\nu\nu'\omega}$ and Eq.~(\ref{equ:defF}), we can now easily obtain $\chi_{d,\text{asym}}^{\nu\nu'\omega}$ as
\begin{equation}
\label{equ:chihighfreq}
 \chi_{r,\text{asym}}^{\nu\nu'\omega}=\chi_{0,r}^{\nu\nu'\omega}-\frac{1}{\beta^2}\sum_{\nu_1\nu_2}\chi_{0,r}^{\nu\nu_1\omega}F_{r,\text{asym}}^{\nu_1\nu_2\omega}\chi_{0,r}^{\nu_2\nu'\omega},
\end{equation}
which completes our analysis of the high-frequency asymptotic behavior of two-particle correlation functions.

\section{Numerical implementations}
\label{sec:numerical_impl}

\begin{figure*}[t!]
	\subfloat{\includegraphics[width=1.0\textwidth]{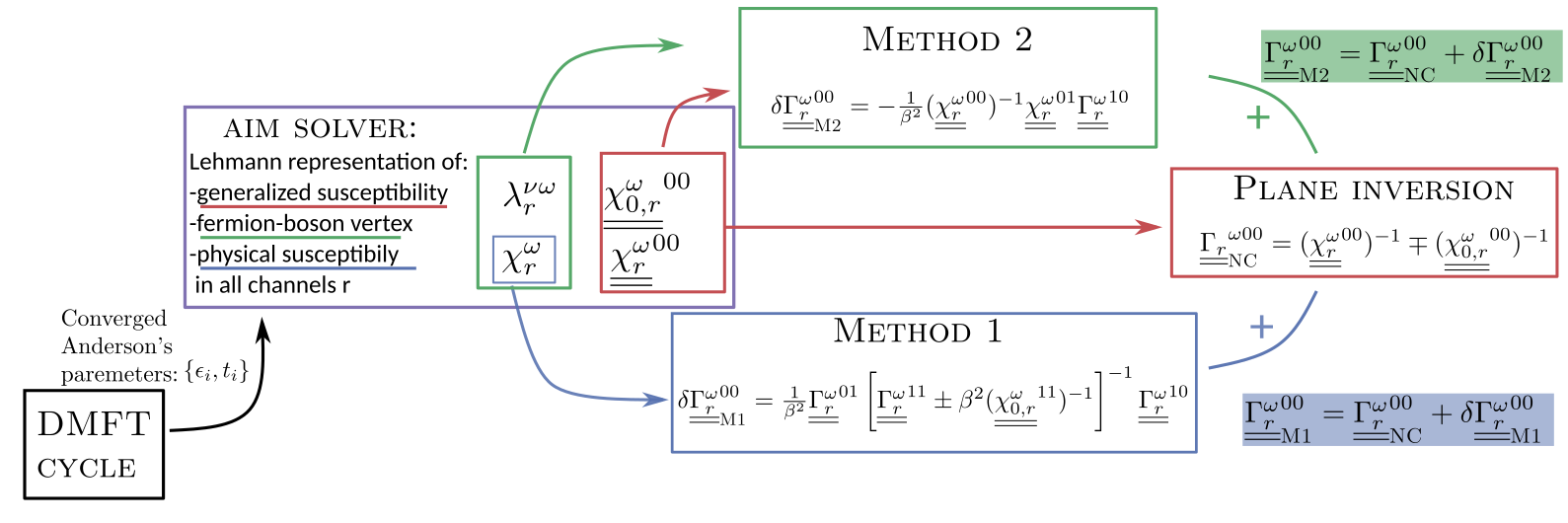}}
	\caption{[Color online] Schematic representation of the proposed procedures to properly invert the BS equations by means of the two correction schemes. From the impurity solver one has access to the asymptotic functions $\chi_r^{\omega}$ and $\lambda_r^{\nu, \omega}$, which are used to construct the correction terms (method 1 in blue and method 2 in green) to the plane inversion of the generalized susceptibility (red). The results for the two methods as discussed in Sec. ~\ref{sec:num_results} converge quickly for increasing low frequency interval $I_0$.}
	\label{fig:flowchart}
\end{figure*}

In order to calculate $\dbunderline{\Gamma_r^{\omega}}$ according to Eq.~(\ref{equ:calcGamma}) one has to invert the two dimensional infinite matrix $\dbunderline{\chi_r^{\omega}}$. In practice, $\chi_r^{\nu\nu'\omega}$, as obtained from a DMFT impurity solver such as exact diagonalization (ED) or quantum Monte Carlo (QMC), is available only on a finite frequency grid with a rather limited number of frequencies. In fact, for a given bosonic frequency $\omega$, the numerical cost for obtaining $\chi_r^{\nu\nu'\omega}$ numerically exactly with any of the current state-of-the-art impurity solvers (ED, QMC, etc..) is typically at least proportional to $\sim\! N^2$, where $N$ is the number of fermionic frequencies. Hence, one has to restrict oneself to a rather small number of frequencies for performing the inversion of Eq.~(\ref{equ:calcGamma}), which might introduce a non-negligible truncation error in the results for $\Gamma_r^{\nu\nu'\omega}$. On the other hand, for large values of the frequencies $\nu$ and $\nu'$, the functions $\chi_r^{\nu\nu'\omega}$ and $\Gamma_r^{\nu\nu'\omega}$ can be replaced by their asymptotic forms~(\ref{equ:chihighfreq}) and (\ref{equ:Gammahighfreq}), respectively. The latter are given in terms of the physical susceptibilities $\chi_r^\omega$ and the fermion-boson vertex $\lambda_r^{\nu\omega}$ which depend on a {\sl single} (fermionic or bosonic) frequency argument (for a fixed value of $\omega$). Hence, the cost of calculating them by means of the impurity solver grows only linearly with the number of frequencies and therefore they can be obtained for a much larger frequency grid (see Appendix~\ref{app:calckernelfunc} for technical details on the calculation of $\chi_r^\omega$ and $\lambda_r^{\nu\omega}$ within ED).

The above discussion suggests the following procedure to determine $\Gamma_r^{\nu\nu'\omega}$ from $\chi_r^{\nu\nu\omega}$: we start by specifying the frequency interval $I=[M_{\text{min}},M_{\text{max}}]$, with $M_{\text{min}},M_{\text{max}}\!\in\!\mathds{Z}$ (for the fermionic Matsubara indices $n$ and $n'$, i.e., $n,n'\!\in\!I$), for which these functions should be computed. We then split it into a (small) low and (large) high frequency part $I_0=[N_{\text{min}},N_{\text{max}}]$, with $M_{\text{min}}\!<\!N_{\text{min}}\!<\!N_{\text{max}}\!<\!M_{\text{max}}\!\in\mathds{Z}$, and $I_1=I \backslash I_0$, respectively (for an illustration see Fig.~\ref{fig:intervals}). Following Ref.~[\onlinecite{Kunes2011}] we can rewrite the matrices $\dbunderline{\chi_r^\omega}$ and $\dbunderline{\Gamma_r^\omega}$ as
\begin{equation}
\label{equ:blockrepgammachi}
 \dbunderline{\chi_r^\omega}=\begin{pmatrix}
	 \dbunderline{\chi_{r}^\omega}^{00} & \dbunderline{\chi_{r}^\omega}^{01} \\ \dbunderline{\chi_{r}^\omega}^{10} &\dbunderline{\chi_{r}^\omega}^{11}
 \end{pmatrix},\quad
\dbunderline{\Gamma_r^\omega}=
\begin{pmatrix}
	 \dbunderline{\Gamma_{r}^\omega}^{00} & \dbunderline{\Gamma_{r}^\omega}^{01} \\ \dbunderline{\Gamma_{r}^\omega}^{10} &\dbunderline{\Gamma_{r}^\omega}^{11}
 \end{pmatrix},
\end{equation}
where the $00$ block contains the values of the respective function for $\nu,\nu'\!\in\!I_0$, the $01$ block for $\nu\!\in\!I_0$ and $\nu'\!\in\!I_1$, the $10$ block for $\nu\!\in\!I_1$ and $\nu'\!\in\!I_0$, and the $11$ block for both $\nu,\nu'\!\in\!I_1$. In the blocks $10$, $01$, and $11$ (i.e., where at least one of the frequencies $\nu$ or $\nu'$ is in the region $I_1$ and, hence, ``large'') we can then replace the exact values for $\chi_r^{\nu\nu'\omega}$ and $\Gamma_r^{\nu\nu'\omega}$ by their asymptotic functions given in Eqs.~(\ref{equ:chihighfreq}) and (\ref{equ:Gammahighfreq}), limiting the numerical treatment of the full frequency dependence to the $00$ region, i.e., for low frequencies $\nu,\nu'\!\in\!I_0$. The goal is then to compute the low-frequency part of $\Gamma_r^{\nu\nu'\omega}$, i.e., $\dbunderline{\Gamma_r^\omega}^{00}$ from Eq.~(\ref{equ:calcGamma}).

An important question concerns the choice of the intervals $I_0$ and $I$. As detailed in Ref.~[\onlinecite{Rohringer2012}], the main low-energy structures that are not captured by the asymptotic functions $\Gamma_{r,\text{asym}}$ and $\chi_{r,\text{asym}}$ arise in a frequency box $I_{\rm low-energy}$ spanned by\footnote{Following the discussion in Ref.~[\onlinecite{Rohringer2012}], the position of the main asymptotic structures of a purely local vertex function in the $\nu$-$\nu'$ frequency space is determined by the maxima of the physical susceptibilities $\chi_d(\nu-\nu')$, $\chi_m(\nu-\nu')$ and $\chi_{pp,\uparrow\downarrow}(\nu+\nu'+\omega)$ (for the $ph$ channels $d$ and $m$). The latter {\sl always} take their largest value at zero frequency, i.e., for the static limit. This leads indeed to the condition $\nu,\nu'=-\omega/2$ for the $ph$ channels $d$ and $m$. Analogous arguments apply to the $pp$ channels $s$ and $t$.  Let us, however, mention that the inclusion of non-local correlations could induce a broadening \cite{Kinza2013a} of the frequency structures, which might require a slightly different choice of the frequency interval.} the corners $(0,0)$, $(0,\pm \omega)$, $(\pm \omega, 0)$, and $(\pm\omega,\pm \omega)$ (with $-$ for $r\!=\!d,m$ and $+$ for $r\!=\!s,t$). In order to take them into account exactly, the inner frequency interval $I_0$ has to be larger than $\omega$. For numerical convenience, it is advantageous to choose $I_0$ and $I$ symmetrically around $I_{\rm low-energy}$.

If we now multiply Eq.~(\ref{equ:calcGamma}) with $\dbunderline{\chi_r^\omega}$ from the left, we obtain a matrix equation for $\dbunderline{\Gamma_r^\omega}$. Using the block representation of Eq.~(\ref{equ:blockrepgammachi}) to separate $\dbunderline{\chi_r^\omega}$ and $\dbunderline{\Gamma_r^\omega}$ into low- and high-frequency contributions, we get
\begin{align}
\label{equ:BSblockequ}
 \frac{1}{\beta^2}
 \begin{pmatrix}
	 \dbunderline{\chi_r^\omega}^{00} & \dbunderline{\chi_r^\omega}^{01} \\
	 \dbunderline{\chi_r^\omega}^{10} & \dbunderline{\chi_r^\omega}^{11}
 \end{pmatrix}&\cdot
\begin{pmatrix}
	 \dbunderline{\Gamma_r^\omega}^{00} & \dbunderline{\Gamma_r^\omega}^{01} \\
	 \dbunderline{\Gamma_r^\omega}^{10} & \dbunderline{\Gamma_r^\omega}^{11}
	\end{pmatrix}=
	\begin{pmatrix}
		\dbunderline{\mathds{1}} & \dbunderline{\mathbb{0}} \\
		\dbunderline{\mathbb{0}} & \dbunderline{\mathds{1}}
 \end{pmatrix}\nonumber\\&\hspace{-1cm}
\mp
\begin{pmatrix}
	 \dbunderline{\chi_r^\omega}^{00} & \dbunderline{\chi_r^\omega}^{01} \\
	 \dbunderline{\chi_r^\omega}^{10} & \dbunderline{\chi_r^\omega}^{11}
 \end{pmatrix}\cdot
\begin{pmatrix}
	 \dbunderline{\chi_{r,0}^\omega}^{00} & \dbunderline{\mathbb{0}} \\
	 \dbunderline{\mathbb{0}} & \dbunderline{\chi_{r,0}^\omega}^{11}
 \end{pmatrix}^{-1},
\end{align}
which leads to four coupled equations for the different blocks\cite{Kunes2011,Hummel2014}. Here we report the first two:
\begin{subequations}
\label{equ:BSblockequseperate}
\begin{align}
\label{equ:BSblockequseperate1}
&\frac{1}{\beta^2}\left[\dbunderline{\chi_r^{\omega}}^{00}\dbunderline{\Gamma_r^{\omega}}^{00}+\dbunderline{\chi_r^{\omega}}^{01}\dbunderline{\Gamma_r^{\omega}}^{10}\right]=\mathds{1}\mp\dbunderline{\chi_r^\omega}^{00}\left(\dbunderline{\chi_{0,r}^\omega}^{00}\right)^{-1},\\
\label{equ:BSblockequseperate2}
&\frac{1}{\beta^2}\left[\dbunderline{\chi_r^{\omega}}^{00}\dbunderline{\Gamma_r^{\omega}}^{01}+\dbunderline{\chi_r^{\omega}}^{01}\dbunderline{\Gamma_r^{\omega}}^{11}\right]=\mp\dbunderline{\chi_r^\omega}^{01}\left(\dbunderline{\chi_{0,r}^\omega}^{11}\right)^{-1},
\end{align}
\end{subequations}
since they are the only ones needed for the derivation of our methods. We note that in these equations all quantities can be extracted directly from the impurity solver (note that $\dbunderline{\Gamma_r^\omega}^{01}$, $\dbunderline{\Gamma_r^\omega}^{10}$ and $\dbunderline{\Gamma_r^\omega}^{11}$ are replaced by their asymptotic functions), except for $\dbunderline{\Gamma_r^\omega}^{00}$, i.e., $\Gamma_r^{\nu\nu'\omega}$ in the low-frequency regime ($\nu,\nu'\!\in\!I_0$), which should be calculated by means of these relations. In fact, from Eq.~(\ref{equ:BSblockequseperate}) one can derive various schemes to determine $\dbunderline{\Gamma_r^\omega}^{00}$, two of which will be illustrated in the following two subsections.

\subsection{Method 1: $\Gamma$'s asymptotics}
\label{subsec:method_1}

The first method for obtaining $\Gamma_r^{\nu\nu'\omega}$ in the low-frequency regime is based on both Eqs.~(\ref{equ:BSblockequseperate}) and uses only the high-frequency asymptotic functions for $\Gamma_r^{\nu\nu'\omega}$. It was first put forward by J. Kune\v{s} in Ref.~[\onlinecite{Kunes2011}] for the particle-hole channels ($r\!=\!d,m$) at $\omega\!=\!0$ only. In this work, the high-frequency behavior of the irreducible vertex was derived by a functional derivative of the self-energy, which leads to the aforementioned restrictions. Our diagrammatic analysis of the vertex asymptotics instead allows for a general formulation including the particle-particle channels and finite frequencies.

Let us briefly recall how this approach can be derived\cite{Kunes2011}: From Eq.~(\ref{equ:BSblockequseperate1}) one obtains $\dbunderline{\Gamma_r^\omega}^{00}$ by applying the inverse of $\dbunderline{\chi_r^\omega}^{00}$ on both sides of the equation. In order to get rid of the asymptotic function $\dbunderline{\chi_r^\omega}^{01}$, one uses Eq.~(\ref{equ:BSblockequseperate2}) which can be recasted into
\begin{equation}
\label{equ:BSexprchiasympt}
 \frac{1}{\beta^2}\dbunderline{\chi_r^\omega}^{01}=-\dbunderline{\chi_r^\omega}^{00}\dbunderline{\Gamma_r^\omega}^{01}\left[\dbunderline{\Gamma_r^\omega}^{11}\pm\beta^2(\dbunderline{\chi_{0,r}^\omega}^{11})^{-1}\right]^{-1}.
\end{equation}
Inserting this into Eq.~(\ref{equ:BSblockequseperate1}) yields
\begin{align}
\label{equ:gamma00final1}
\frac{1}{\beta^2}\dbunderline{\Gamma_r^\omega}^{00}&=(\dbunderline{\chi_r^\omega}^{00})^{-1}\mp(\dbunderline{\chi_{0,r}^\omega}^{00})^{-1}\nonumber\\+\frac{1}{\beta^2}&\dbunderline{\Gamma_r^\omega}^{01}\left[\dbunderline{\Gamma_r^\omega}^{11}\pm\beta^2(\dbunderline{\chi_{0,r}^\omega}^{11})^{-1}\right]^{-1}\dbunderline{\Gamma_r^\omega}^{10}.
\end{align}
The right-hand side (r.h.s.) of this equation can be interpreted straightforwardly: the first line corresponds to the calculation of $\Gamma_r^{\nu\nu'\omega}$ in the low-frequency regime (without higher frequencies), while the second line represents a correction due to the high-frequency asymptotic contributions.
We note that the calculation of this correction term requires also an additional inversion of the high-frequency parts of $\Gamma_r^{\nu\nu'\omega}$ and $\chi_r^{\nu\nu'\omega}$ (i.e., of the term in the square brackets). However, the latter can be obtained at much lower cost compared to the full vertex function and, hence, they are available for a much larger frequency grid which reduces the error of this inversion. Moreover, the advantage of the approach is that it does not require the calculation of the fermion boson couplings $\lambda_r^{\nu \omega}$ but only the determination of $\chi_r^{\omega}$, which is numerically significantly less demanding.

\subsection{Method 2: $F$'s asymptotics}
\label{subsec:method_2}

The second method for calculating $\Gamma_r^{\nu\nu'\omega}$ in the low-frequency regime uses {\sl exclusively} Eq.~(\ref{equ:BSblockequseperate1}). From this, one easily obtains
\begin{align}
\label{equ:gamma00final2}
 \frac{1}{\beta^2}\dbunderline{\Gamma_r^\omega}^{00}=(\dbunderline{\chi_r^\omega}^{00})^{-1}&\mp(\dbunderline{\chi_{0,r}^\omega}^{00})^{-1}\nonumber\\-&\frac{1}{\beta^2}(\dbunderline{\chi_r^\omega}^{00})^{-1}\dbunderline{\chi_r^\omega}^{01}\dbunderline{\Gamma_r^\omega}^{10}.
\end{align}
Similar to method $1$, the first line corresponds to the calculation of $\Gamma_r^{\nu\nu'\omega}$ in the low-frequency regime, without the higher frequencies, while the term in the second line represents a correction from the high-frequency asymptotic contributions. The advantage of this approach w.r.t. the one described in the previous section is that it does not require an additional inversion of the high-frequency asymptotic contributions. On the other hand, it makes necessary the determination of the asymptotic functions for $\chi_r^{\nu\nu'\omega}$ (and not only for $\Gamma_r^{\nu\nu'\omega}$), including the evaluation of $\lambda_r^{\nu\omega}$. For a computation within DMFT this represents, as discussed above and detailed in Appendix~\ref{app:calckernelfunc}, not a real obstacle since these functions can be obtained relatively easily from the impurity solver.

\begin{figure}[t!]
	\subfloat{\includegraphics[width=1.0\textwidth]{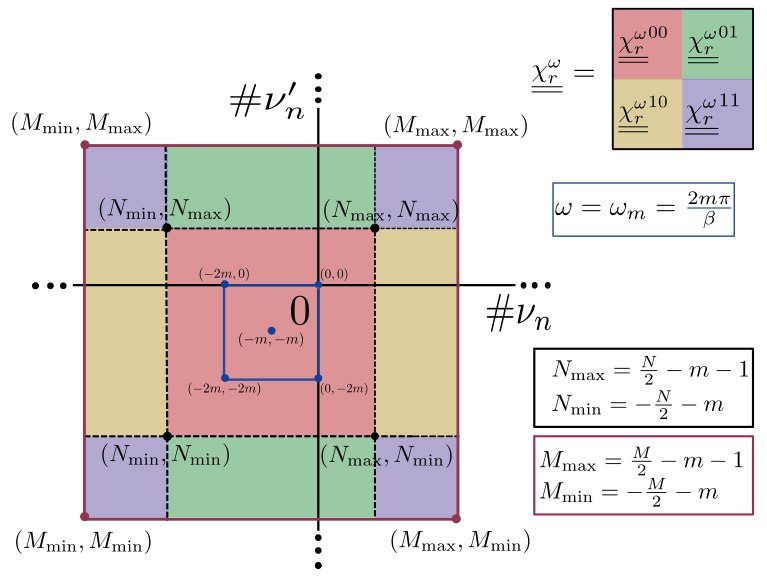}}
	\caption{[Color online] Visual illustration of the block construction (see Sec.~\ref{sec:numerical_impl}).}
	\label{fig:intervals}
\end{figure}

\section{Numerical results}
\label{sec:num_results}

\begin{figure*}[t!]
\centering
\subfloat{\includegraphics[width=0.50\textwidth]{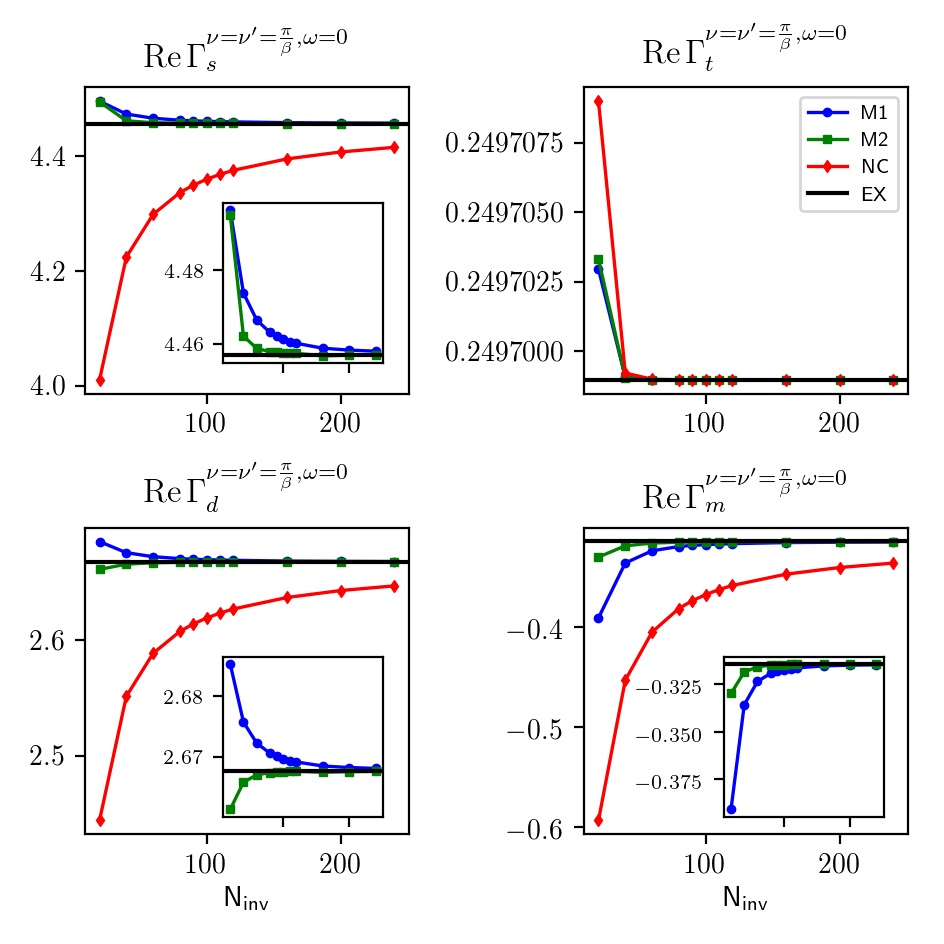}}
\subfloat{\includegraphics[width=0.50\textwidth]{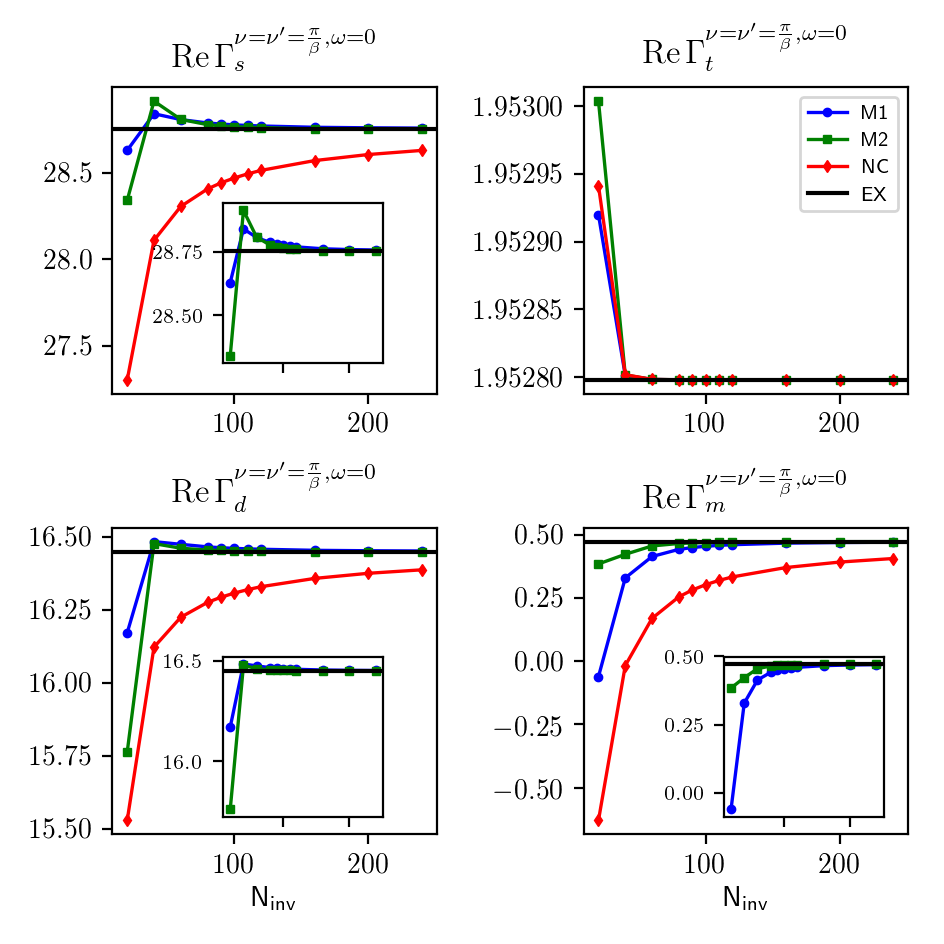}}
	\caption{[Color online] $\Gamma_r^{\nu\nu'\omega}$ for $r=\lbrace s,t,d,m \rbrace$, evaluated at $\nu = \nu' = \pi/\beta$ and $\omega=0$ as a function of the inversion range $\text{N}_{\text{inv}}$, for $U=1$ (left panels) and $U=1.75$ (right panels). Uncorrected results [red, first lines of Eqs.~(\ref{equ:gamma00final1}) and (\ref{equ:gamma00final2})] are compared to results corrected by method 1 [blue, second line of Eqs.~(\ref{equ:gamma00final1})], method 2 [green, second line of Eqs.~(\ref{equ:gamma00final2})] and the exact result (black). The insets provide a comparison between only the two correction methods on a smaller scale.}
	\label{fig:comparison_EXACTINV_M1_M2_Om0}
\end{figure*}

In this section, we present our results for $\Gamma_r^{\nu\nu'\omega}$ obtained by the methods discussed in the previous Secs.~\ref{subsec:method_1} and \ref{subsec:method_2}. For the sake of clarity, we will focus on the DMFT solution of the half-filled Hubbard model in $3d$ [see Eq.~(\ref{eq:defhamilt})] for which a comparison with numerically exact results is possible. Specifically, we consider the two values of the Hubbard interaction $U\!=\!1D$ and $U\!=\!1.75D$, respectively, at a temperature $T\!=\!0.02D$ with $D\!=\!2\sqrt{6}t$ (which correspond to twice the standard deviation of the $3d$ non-interacting density of states). The selected interaction strengths correspond to weak coupling ($U\!=\!1$) and to the highly relevant intermediate-to-strong coupling regime near the Mott MIT ($U\!=\!1.75$), respectively, which allow for a representative benchmark of our newly developed techniques.

For evaluating the performance of the new methods we have pursued the following strategy, which is illustrated in the flowchart in Fig.~\ref{fig:flowchart} : We calculate the irreducible vertex $\Gamma_r^{\nu\nu'\omega}$ (for selected values of $\omega$) in a (small) interval $I_0\!=\![N_{\text{min}},N_{\text{max}}]$ of fermionic frequencies (see Sec.~\ref{sec:numerical_impl}) by means of Eq.~(\ref{equ:calcGamma}) without any corrections, and compare it to the values obtained by correcting the latter results with method 1 [Eq.~(\ref{equ:gamma00final1})] and method 2 [Eq.~(\ref{equ:gamma00final2})] (here and in the following, all frequency values and intervals refer to the corresponding integer valued Matsubara indices). According to the discussion in Sec.~\ref{sec:numerical_impl}, we typically consider a frequency range for $\nu$ and $\nu'$ for which the main structures of the corresponding two-particle susceptibilities and vertices are centered (see also Ref.~[\onlinecite{Rohringer2012}]), i.e., $N_{\text{max}}\!+\!1\!=\!-N_{\text{min}}\!\mp 2m\!=\!N_{\text{inv}}/2\!\mp\!m$ (for the $ph$ and the $pp$ channels, respectively), where $N_{\text{inv}}$ denotes the total number of frequencies used for the matrix inversion in Eq.~(\ref{equ:calcGamma}) and $m$ is the index of the bosonic Matsubara frequency $\omega$. An illustration of these frequency intervals is given in Fig.~\ref{fig:intervals} (red block). The correction terms are obtained in the (much larger) frequency interval $I\!=\![M_{\text{min}},M_{\text{max}}]$, which has been chosen in such a way that all results presented in the following are converged w.r.t. the size of $I$. Specifically, the $\chi_r^{\omega}$'s have been calculated for $2001$ bosonic ($M_{\text{max}}\!=\!-M_{\text{min}}\!=\!1000$) Matsubara frequencies while the $\lambda_r^{\nu\omega}$'s have been evaluated for $240$ fermionic ($M_{\text{max}}\!+\!1\!=\!-M_{\text{min}}\!\mp\!2m\!=\!120\!\mp\!m$, see Fig.~\ref{fig:intervals}) Matsubara frequencies where again the fermionic frequency interval has been centered around $\mp m$. Finally, the ``exact'' solution for $\Gamma_r^{\nu\nu'\omega}$, which serves as a benchmark for the performance of our new approaches, has been obtained from the plain inversion of Eq.~(\ref{equ:calcGamma}) (i.e., without any corrections) for a very large frequency grid $I_L\!=\![L_{\text{min}},L_{\text{max}}]$ with $L_{\text{max}}\!+1\!\!=\!-L_{\text{min}}\!\mp\!2m\!=\!L_{\text{inv}}/2\!\mp\!m$ with $L_{\text{inv}}\!=\!320$, and results have been extrapolated to $L \rightarrow\infty$.
All calculations for obtaining $\chi_r^{\nu\nu'\omega}$, $\chi_r^\omega$ and $\lambda_r^{\nu\omega}$ have been carried out by means of an ED impurity solver\footnote{For the calculation of the full three-frequency dependent $\chi_r^{\nu\nu'\omega}$ the open source exact diagonalization (full-ED) code {\it pomerol} has been adopted, see Ref.~[\onlinecite{andrey_e_antipov_2015_17900}].} (see Ref.~[\onlinecite{Toschi2007}] and Appendix~\ref{app:calckernelfunc}) whereupon the AIM related to the DMFT solution of (\ref{eq:defhamilt}) has been parametrized by four bath sites.

Figure \ref{fig:comparison_EXACTINV_M1_M2_Om0} shows $\Gamma_r^{\nu\nu'\omega}$ for fixed values of the Matsubara frequencies, i.e., $\nu\!=\!\nu'\!=\!\pi/\beta$, $\omega\!=\!0$, as a function of the total number of fermionic frequencies $N_{\text{inv}}$ which have been used for the inversion of $\chi_r^{\nu\nu'\omega}$ in the $00$ block [see first lines of Eqs.~(\ref{equ:calcGamma}) and (\ref{equ:blockrepgammachi})]. The uncorrected values (red) are compared to the corresponding results of method 1 [blue, Eq.~(\ref{equ:gamma00final1})] and method 2 [green, Eq.~(\ref{equ:gamma00final2})] as well as the exact solution (black), for $U\!=\!1$ (left panels) and $U\!=\!1.75$ (right panels). We can see that even for a relatively large value of $\text{N}_{\text{inv}}\!=\!240$ the non-corrected results substantially deviate from the corrected ones which both converge rapidly to the exact value in all channels and for both values of $U$ (up to a relative error lower than $10^{-3}$).
As expected, the non-corrected results are drastically affected by varying $\text{N}_{\text{inv}}$, with a (maximal) relative error of $\sim 100\%$ for $U=1$ and $\sim 200\%$ for $U=1.75$ w.r.t. the exact solution in the most sensitive magnetic channel for the lowest value of $\text{N}_{\text{inv}}\!=\!20$.

\begin{figure}[t!]
\subfloat{\includegraphics[width=1.0\textwidth]{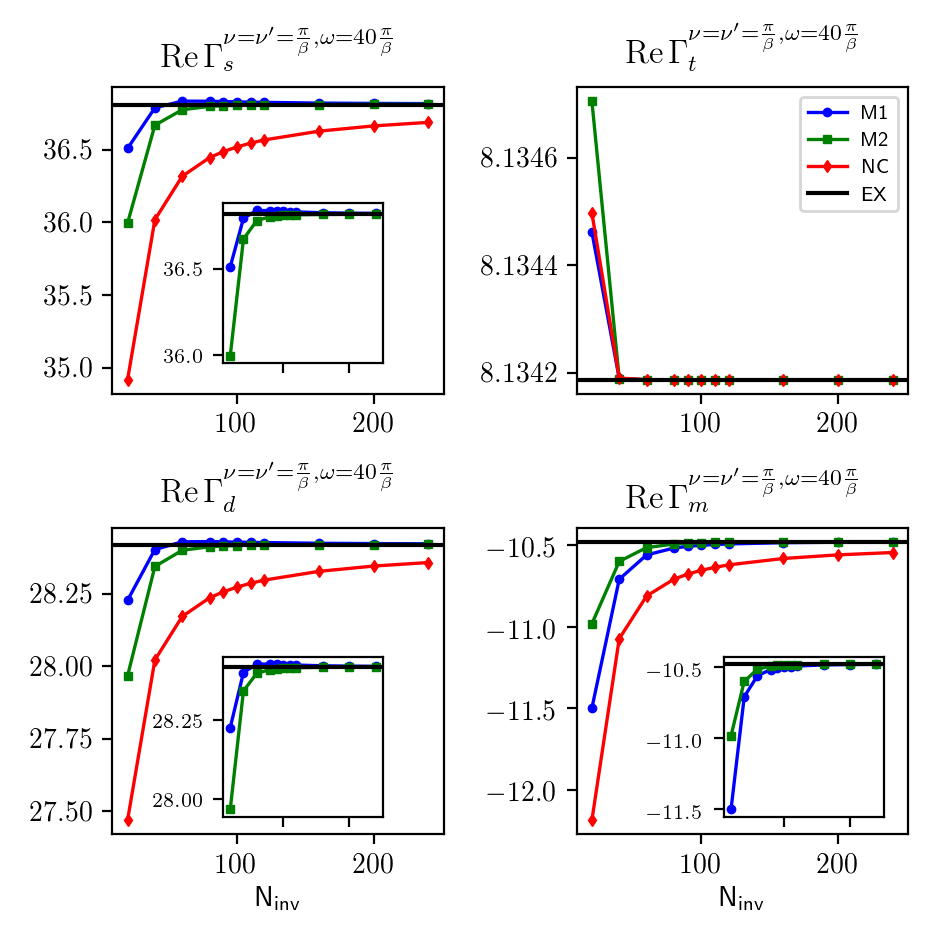}}
\caption{[Color online] Same as in Fig.~\ref{fig:comparison_EXACTINV_M1_M2_Om0} but for $\omega\!=\!40\pi/\beta$ and $U\!=\!1.75$ only.}
\label{fig:comparison_EXACTINV_M1_M2_Om20}
\end{figure}

\begin{figure*}[t!]
\begin{center}
\subfloat{\includegraphics[width=0.5\textwidth]{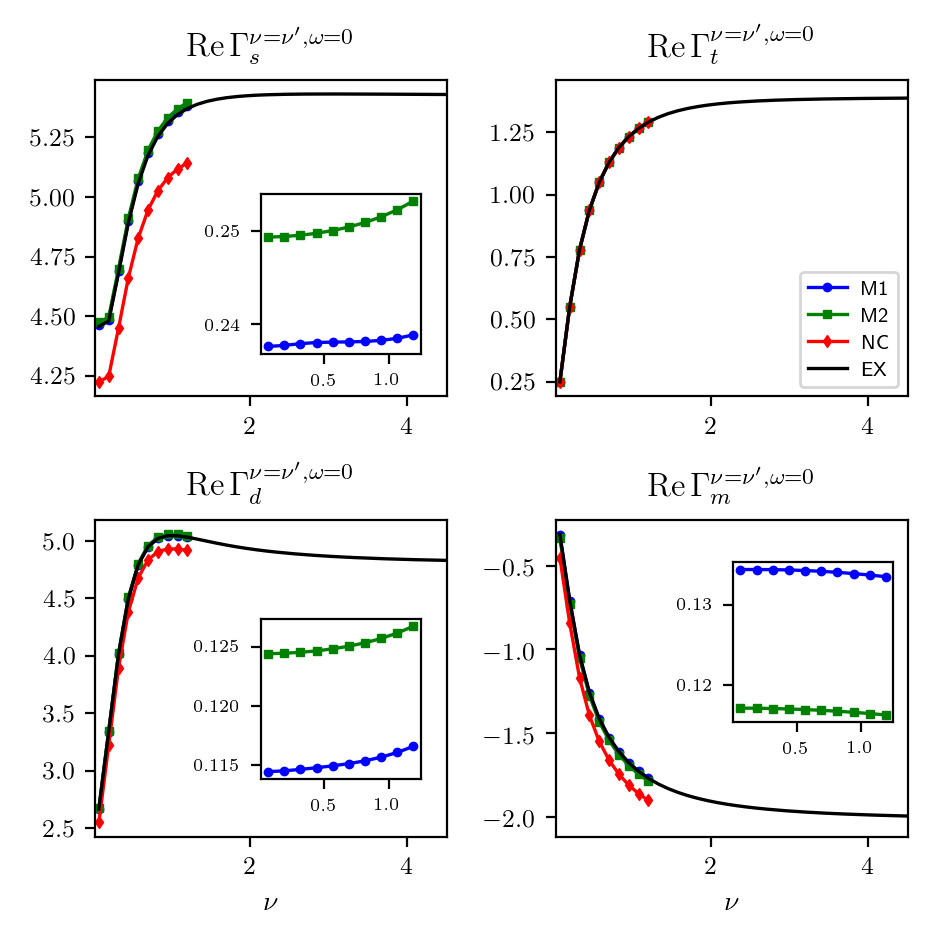}}
\subfloat{\includegraphics[width=0.5\textwidth]{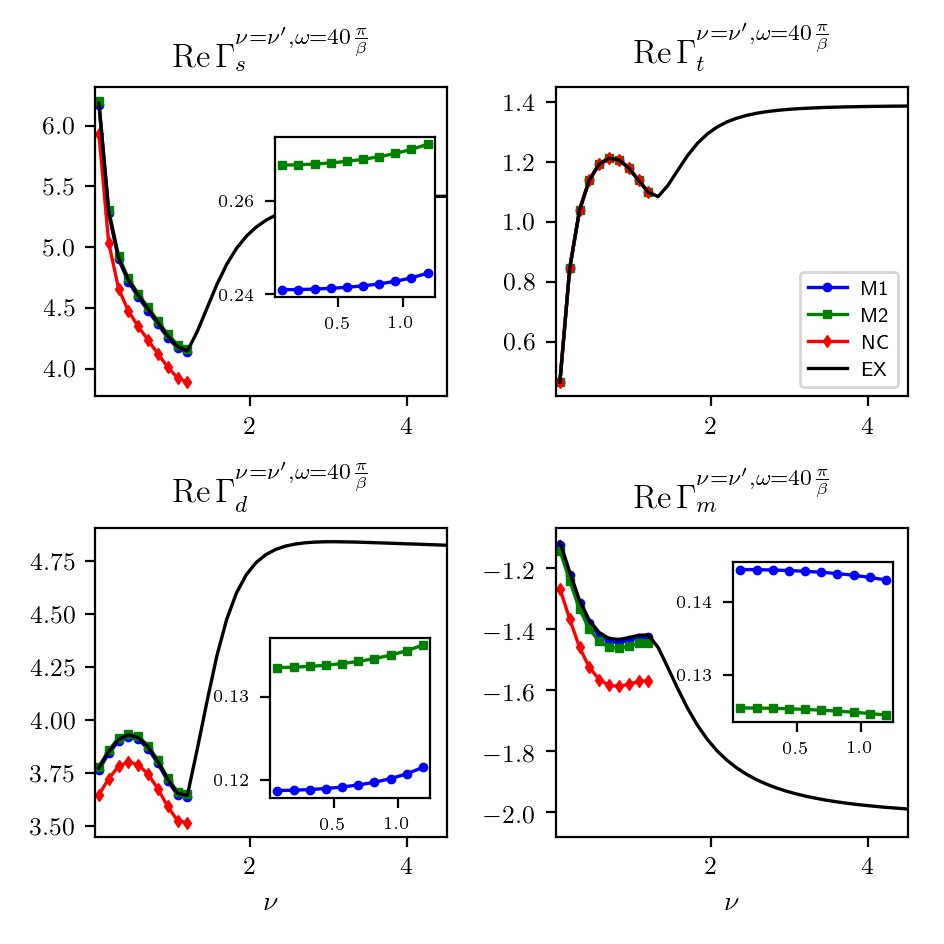}}
\caption{[Color online]  $\Gamma_r^{\nu\nu'\omega}$ for $r=\lbrace s,t,d,m \rbrace$, evaluated along the diagonal $\nu = \nu'$ for $U=1$, $\omega=0$ (left panels) and $\omega=40\pi/\beta$ (right panels). Results of method $1$ (blue) and method $2$ (green) are compared to a plain inversion (red) of Eq.~(\ref{equ:calcGamma}) for $N_{\text{inv}}\!=\!40$ and the exact solution (black). Insets show the correction terms only.}
\label{fig:comparison_gamma_U1}
\end{center}
\end{figure*}

\begin{table*}[t!]
\begin{center}
\setlength\doublerulesep{1.5pt}
\begin{adjustbox}{max width=0.32\textwidth}
\begin{tabular}{|c|cc|cc|}
\hline
$N_{\text{inv}}$ 	& $\delta \Gamma^{\text{M1}}_{s}(\omega=0)$ & $\delta \Gamma^{\text{M1}}_{s}(\omega=40 \frac{\pi}{\beta})$ & $\delta \Gamma^{\text{M2}}_{s}(\omega=0)$ & $\delta \Gamma^{\text{M2}}_{s}(\omega=40 \frac{\pi}{\beta})$ \\
\hline
$40$		& -0.256  &   -0.276   &     -0.238 &  -0.246 \\
$80$		&  -0.126  & -0.129   &     -0.121 &  -0.122 \\
$120$ 		& -0.0842  &   -0.0850   &     -0.0814&  -0.0818  \\
$160$		&  -0.0631 & -0.0634    &    -0.061 &  -0.0613\\
$200$		& -0.0505  &  -0.0507   &    -0.0492 &  -0.0493	\\
$240$ 		& -0.0421  &   -0.0422  &    -0.0412 &   -0.04127\\ \hhline{=====}
$40$		&  -0.759 &  -0.801    &    -0.706 & -0.699  \\
$80$		& -0.382  & -0.389  &  -0.357  &  -0.357 \\
$120$ 		&  -0.257 &   -0.259   &     -0.243 &  -0.243  \\
$160$		& -0.193  &   -0.193  &  -0.183 & -0.184 \\
$200$		&  -0.154 &  -0.155   &   -0.148 &  -0.148	\\
$240$ 		& -0.129 &    -0.129 &    -0.124 &  -0.124 \\
\hline
\end{tabular}
\end{adjustbox}
\begin{adjustbox}{max width=0.32\textwidth}
\begin{tabular}{|c|cc|cc|}
\hline
$N_{\text{inv}}$ 	& $\delta \Gamma^{\text{M1}}_{d}(\omega=0)$ & $\delta \Gamma^{\text{M1}}_{d}(\omega=40 \frac{\pi}{\beta})$ & $\delta \Gamma^{\text{M2}}_{d}(\omega=0)$ & $\delta \Gamma^{\text{M2}}_{d}(\omega=40 \frac{\pi}{\beta})$ \\
\hline
$40$		& -0.128  &   -0.137   &    -0.117 &  -0.123 \\
$80$		&  -0.0631 &  -0.0643  &     -0.0598 &  -0.0607 \\
$120$ 		& -0.0421  &   -0.0425   &     -0.0406 &  -0.0408  \\
$160$		& -0.0315  &  -0.0317  &      -0.0305 &  -0.0307\\
$200$		& -0.0253  &   -0.0254  &     -0.0246 &  -0.0247	\\
$240$ 		& -0.0211  & -0.0211   &  -0.0206  &  -0.0206  \\ \hhline{=====}
$40$		& -0.374 &  -0.395   &    -0.344 & -0.353  \\
$80$		&  -0.190 &  -0.194  & -0.176  &  -0.177 \\
$120$ 		& -0.128  &   -0.129   &   -0.121 &   -0.121 \\
$160$		& -0.0962 & -0.0966  &    -0.0914 & -0.0918 \\
$200$		&  -0.0772 &  -0.0774  &   -0.0739  &  -0.0742	\\
$240$ 		&  -0.0644 & -0.0646 &     -0.0621 &  -0.0622 \\
\hline
\end{tabular}
\end{adjustbox}
\begin{adjustbox}{max width=0.32\textwidth}
\begin{tabular}{|c|cc|cc|}
\hline
$N_{\text{inv}}$ 	& $\delta \Gamma^{\text{M1}}_{m}(\omega=0)$ & $\delta \Gamma^{\text{M1}}_{m}(\omega=40 \frac{\pi}{\beta})$ & $\delta \Gamma^{\text{M2}}_{m}(\omega=0)$ & $\delta \Gamma^{\text{M2}}_{m}(\omega=40 \frac{\pi}{\beta})$ \\
\hline
$40$		& -0.116  &  -0.124   &     -0.133 &  -0.142 \\
$80$		&  -0.0619  & -0.0631  &     -0.0662 &  -0.0673 \\
$120$ 		&  -0.0418 &   -0.0422    & -0.0437     &   -0.0440 \\
$160$		& -0.0316  &  -0.0317  &     -0.0325 & -0.0327 \\
$200$		& -0.0253  &    -0.0254 &    -0.0259 &  -0.0259	\\
$240$ 		&  -0.0211 &  -0.0211 &     -0.0215 &  -0.0215 \\ \hhline{=====}
$40$		& -0.345 &  -0.364  &     -0.436 & -0.456 \\
$80$		&  -0.188 &  -0.191  &    -0.211 &  -0.214 \\
$120$ 		&  -0.127 &    -0.128   &     -0.138 &   -0.139 \\
$160$		&  -0.0964 & -0.0969  &    -0.102 & -0.103 \\
$200$		& -0.0773  &  -0.0775   &    -0.0808 &  -0.0810	\\
$240$ 		& -0.0645  &  -0.0646 &  -0.0668 & -0.0670 \\
\hline
\end{tabular}
\end{adjustbox}
\caption{Differences $\delta \Gamma^{\eta}_{r}(\omega)$, with $\eta\!=\!\text{M1}$ for method $1$ and $\eta\!=\!\text{M2}$ for method $2$, between the non-corrected and the corrected results, as provided by the two different methods. The data are reported for different sizes of the frequency box $N_{\text{inv}}$ used for the inversion, and for $\omega\!=\!0$ and $\omega\!=\!40\frac{\pi}{\beta}$. The upper panel refers to $U=1$, and the lower to $U=1.75$.}
\label{tab:table_U1}
\end{center}
\end{table*}

\begin{figure*}[t!]
\begin{center}
\subfloat{\includegraphics[width=0.5\textwidth]{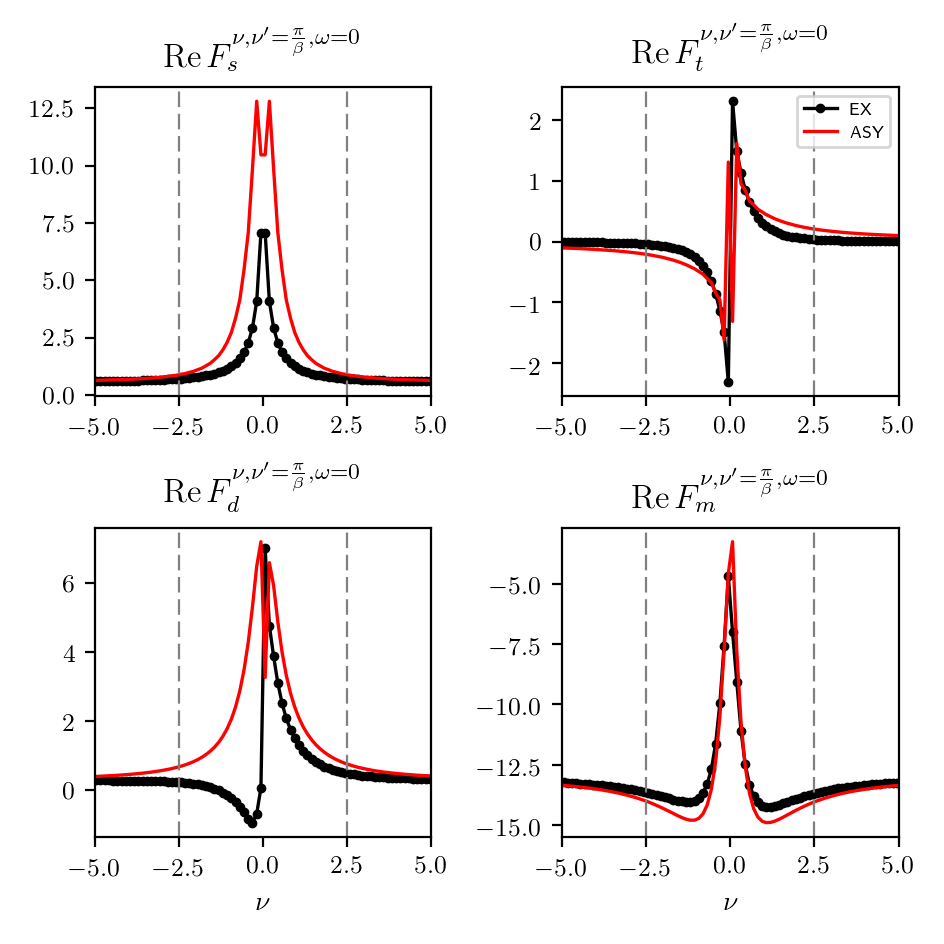}}
\subfloat{\includegraphics[width=0.5\textwidth]{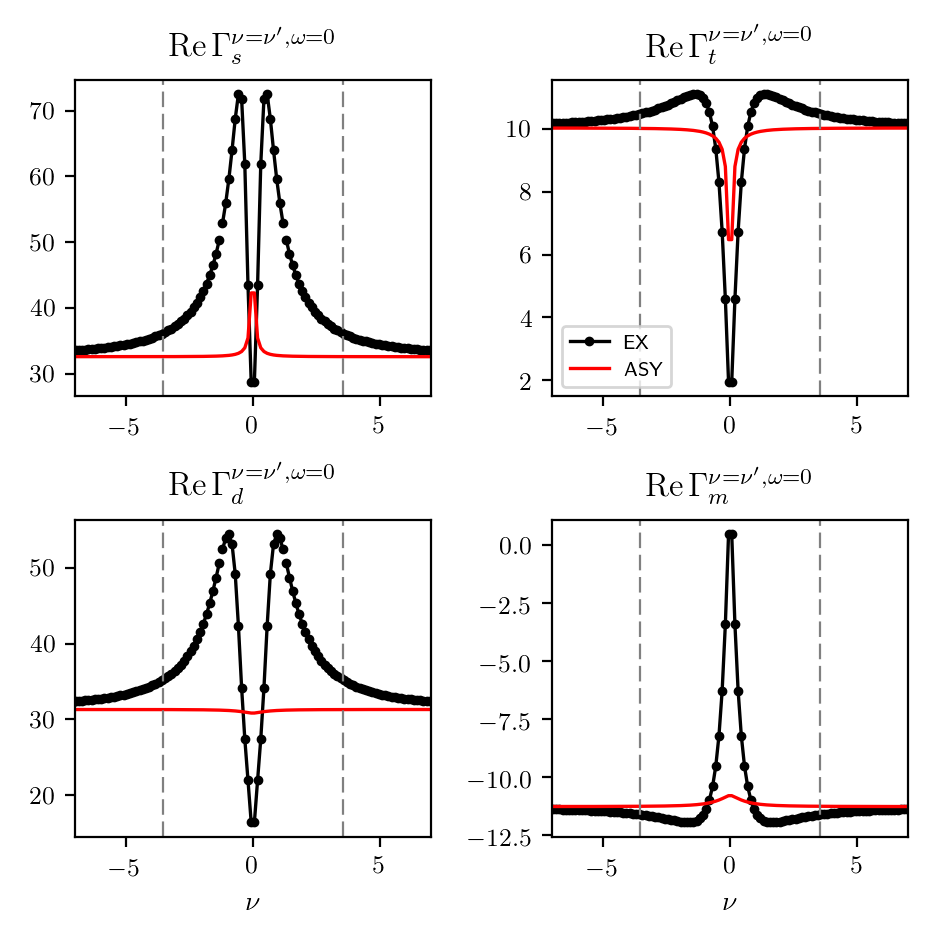}}
\caption{[Color online] Left panel: (ED)DMFT full vertex $F_r^{\nu\nu'\omega}$ (black), with $r=\lbrace s,t,d,m \rbrace$ as a function of $\nu$ for $\nu'=\pi/\beta$, $\omega=0$ and $U=1.75$. Right panel: "Exact" (see definition in the text) $\Gamma_r^{\nu\nu'\omega}$ (black), with $r=\lbrace s,t,d,m \rbrace$  evaluated for $\nu=\nu'$, $\omega=0$ and  $U=1.75$. In both panels the red line shows the behavior of the respective asymptotic functions [see Eq.~\eqref{eq:f_eqs_asy}]. The dashed lines mark the edges of the $(00)$ block for $N_{\text{inv}}\!=\!40$.}
\label{fig:vertex}
\end{center}
\end{figure*}

In Fig.~\ref{fig:comparison_EXACTINV_M1_M2_Om20}, we report an analogous analysis as in Fig.~\ref{fig:comparison_EXACTINV_M1_M2_Om0}, but for a finite value of $\omega\!=\!40\pi/\beta$. The overall picture is very similar as for $\omega\!=\!0$ which confirms the applicability of the proposed methods for a finite bosonic transfer frequency. Interestingly, for both values of $\omega$ the triplet channel is not affected by the correction terms and converges rapidly to the exact solution. This can be attributed to the fact that for the triplet vertices $\Gamma_t^{\nu\nu'\omega}$ and $F_t^{\nu\nu'\omega}$, respectively, the constant background proportional to $U$ is absent. Hence, in their asymptotic high-frequency regime the latter are dominated by the two diagonal structures\cite{Rohringer2012,Rohringer2014} at $\nu\!=\!\nu'$ and $\nu\!=\!\omega\!-\!\nu'$, which originate from $\chi_{d/m}^{\nu'-\nu}$ and $\chi_{d/m}^{\omega-\nu-\nu'}$ [see Eqs.~(\ref{equ:Gammathighfreq}) and (\ref{eq:ft_eqs_asy})]. Consequently, the actual inversion of the corresponding $\chi_t^{\nu\nu'\omega}$ depends only very weakly on the size of $N_{\text{inv}}$. More specifically, one can infer the (almost) vanishing of the correction terms for $\Gamma_t^{\nu\nu'\omega}$ in the second lines of Eqs.~(\ref{equ:gamma00final1}) and (\ref{equ:gamma00final2}) also directly from the observation that the asymptotic contributions $\dbunderline{\Gamma_t^\omega}^{10}$ and $\dbunderline{\Gamma_t^\omega}^{10}$ are very small. In fact, they involve the physical susceptibilities $\chi_{d/m}^\omega$ for $\omega\!\ne\!0$ in Eq.~(\ref{equ:Gammathighfreq}),\footnote{Note that for the $10$ and the $01$ region in the $\nu-\nu'$ frequency plane $\nu\!\ne\!\pm\nu'(+\omega)$.} while the latter are typically taking their maximum at $\omega\!=\!0$ and are rapidly decreasing with increasing $\omega$.

In Fig.~\ref{fig:comparison_gamma_U1}, we turn our attention to the frequency dependence of the correction terms obtained by the two methods. Specifically, we show $\Gamma_r^{\nu\nu'\omega}$ along the diagonal $\nu\!=\!\nu'$ for $\omega\!=\!0$ (left panels) and $\omega\!=\!40\pi/\beta$ (right panels) at $U\!=\!1$ (corresponding results for $U\!=\!1.75$ are given in Appendix~\ref{app:gamma_1.75}) where the uncorrected solution (red) has been obtained for $N_{\text{inv}}\!=\!40$. As expected from the discussion above, the corrections from both methods $1$ (blue) and $2$ (green) are almost equivalent and provide a relevant improvement for the calculations of $\Gamma_r^{\nu\nu'\omega}$ for all channels apart from the triplet one ($r\!=\!t$), for which already the plain inversion yields an almost exact result. Interestingly, the corrections appear to have a weak dependence on ($\nu,\nu'$) as can be seen in the insets in Fig.~\ref{fig:comparison_gamma_U1} which shows only the corrections provided by method $1$ (blue line) and method $2$ (green line). Indeed, one appreciates how the corrections given by both methods vary less then $1 \%$ along the diagonal $\nu = \nu'$ (for the case $U=1.75$ this is no longer true, see Appendix \ref{app:gamma_1.75}). This can be explained by a similar argument as given for the vanishing of the corrections for the triplet vertex: rewriting, for instance, the correction term for method $1$ in the second line of Eq.~(\ref{equ:gamma00final1}) into a more explicit form we obtain (apart from a prefactor $1/\beta^2$)
\begin{equation}
\label{equ:corroneexplicit}
 \sum_{\nu_1,\nu_2}\Gamma_{r,\text{asym}}^{\nu\nu_1\omega}\left[\dbunderline{\Gamma_r^\omega}^{11}\pm\beta^2(\dbunderline{\chi_{0,r}^{\omega11}})^{-1}\right]^{-1}_{\nu_1\nu_2}\Gamma_{r,\text{asym}}^{\nu_2\nu'\omega},
\end{equation}
where $\nu,\nu'\in I_0$ and the sums over $\nu_1$ and $\nu_2$ run over the high-frequency interval $I \backslash I_0$. Clearly, $\nu\!\ne\!\nu_1$ and $\nu'\!\ne\!\nu_2$ and the $\chi_r^{\omega}$'s in the asymptotic expressions for $\Gamma_{r,\text{asym}}^{\nu\nu_1\omega}$ and $\Gamma_{r,\text{asym}}^{\nu_2\nu'\omega}$ in Eqs.~(\ref{equ:Gammahighfreq}), which contribute here only for $\omega\!\ne\!0$, are, hence, typically very small. Consequently, $\Gamma_{r,\text{asym}}^{\nu\nu_1\omega}$ and $\Gamma_{r,\text{asym}}^{\nu_2\nu'\omega}$ can be both approximated by the corresponding first terms in Eqs.~(\ref{equ:Gammahighfreq}) which are just constants proportional to $U$. When we insert this approximation, i.e., $\Gamma_{r,\text{asym}}^{\nu\nu_1\omega}\!=\!\Gamma_{r,\text{asym}}^{\nu_2\nu'\omega}\!\sim\!U$, into the expression for the correction term of method $1$ in Eq.~(\ref{equ:corroneexplicit}), the latter obviously becomes frequency independent. An analogous argument explains also the frequency independence\footnote{For the correction term in second line of Eq.~(\ref{equ:gamma00final2}) we can actually prove only its $\nu'$ independence explicitly. However, for a system with time inversion symmetry, we have that $\Gamma_r^{\nu\nu'\omega}\!=\!\Gamma_r^{\nu'\nu\omega}$ (and the same for the corresponding asymptotic functions) and, consequently, the correction also does not depend on $\nu$.} of the correction term obtained in method $2$ [second line of Eq.~(\ref{equ:gamma00final2})] for which the above considerations also allow for a simplification of the calculation procedure. Considering, e.g., for the density channel, that for the (off-diagonal) $(10)$ frequency region $F_{d,\text{asym}}^{\nu\nu'\omega}\!\sim\!U\!+\!U^2\chi_d^\omega\!+\!U\lambda_d^{\nu\omega}$ [Eq.~(\ref{eq:fd_eqs_asy})], we obtain for the correction term (apart from a prefactor $1/\beta^2$) approximately
\begin{equation}
\label{eq:gamma00_m2_explicit}
\begin{split}
\sum_{\nu_1} \left[\dbunderline{\chi_d^{\omega}}^{00}\right]^{-1}_{\nu\nu_1}&G(\nu_1)G(\nu_1+\omega)[U+U^2 \chi_{d}^{\omega}+U\lambda_{d}^{\nu_1 \omega}]\nonumber\\\times&\sum_{\nu_2} \frac{1}{i\nu_2} \frac{1}{i(\nu_2+\omega)}U.
\end{split}
\end{equation}

where $\left[\dbunderline{\chi_d^{\omega}}^{00}\right]^{-1}$ represents the exact inversion of $\chi_d^{\nu\nu'\omega}$ in the small frequency interval $I_0$ and the (right) outer legs Green's functions of $\chi_d^{\nu_1\nu_2\omega}$, which depend on $\nu_2\!\in\!I \backslash I_0$, have been replaced by their asymptotic values (which allows the corresponding sum over $\nu_2$ to be evaluated analytically).

Given the small dependence of the correction on the fermionic frequencies, one can extract the values of the corrections provided by the two methods at $\nu = \nu' = \pi/\beta$ ($n=n'=0$) in order to give an estimation of the convergence with respect to $\text{N}_{\text{inv}}$. These are reported in Tab.~\ref{tab:table_U1} for two values of $U\!=\!1$ and $U\!=\!1.75$ as well as $\omega\!=\!0$ and $\omega\!=\!40\pi/\beta$ , for all channels but the triplet one, whose corrections can be assumed negligible (see previous discussion). The overall picture that can be obtained from Table \ref{tab:table_U1} confirms the similarity of method $1$ and method $2$ regarding their results and predicts a relatively weak dependence of the corrections on the bosonic Matsubara frequency $\omega$ (as it is the case for the fermionic ones $\nu$ and $\nu'$). For the correction term of method $1$ in the second line of Eq.~(\ref{equ:gamma00final1}) this observation can be ascribed to the fact that for a given bosonic frequency $\omega$ the calculation intervals $I_0$ and $I$ are centered around $\nu,\nu'\!=\!\mp\omega/2$ for the $ph$ ($r\!=\!d,m$) and the $pp$ ($r\!=\!s,t$) channels, respectively. In fact, this centering can be also realized by performing the shift $\nu\!\rightarrow\!\nu\!\mp\!\omega/2$ and $\nu'\!\rightarrow\!\nu\!\mp\!\omega/2$ in $\Gamma_{r,\text{asym}}^{\nu\nu'\omega}$ in Eq.~(\ref{equ:Gammahighfreq}), which renders the latter indeed independent\footnote{The term $\chi_{0,r}^{\nu\nu'\omega}$ in the second line of Eq.~(\ref{equ:gamma00final1}) behaves like $\sim 1/[i\nu(i\nu\!\pm\! i\omega)]$, which, hence, exhibits only a very weak $\omega$ dependence if the interval $I_0$ is chosen considerably larger than $\omega$, i.e., if $\nu,\nu'\!\gg\!\omega$.} of $\omega$. Although the mechanism responsible for the $\omega$ independence of the correction in method $2$ is less transparent [$F_{r,\text{asym}}^{\nu\nu'\omega}$ explicitly depends on $\omega$, see Eq.~\ref{eq:f_eqs_asy}], it could rely on a mutual compensation of $\omega$-dependent terms in the second line of Eq.~\eqref{equ:gamma00final2}, namely, $(\dbunderline{\chi_r^\omega}^{00})^{-1}$ and $\dbunderline{\chi_r^\omega}^{01}$.

Let us finally investigate the frequency structure of the asymptotic functions themselves. In Fig.~\ref{fig:vertex}, we compare $\Gamma_{r,\text{asym}}^{\nu\nu'\omega}$ (right panels) and $F_{r,\text{asym}}^{\nu\nu'\omega}$ (left panels) to the corresponding exact values for $\omega\!=\!0$ at $U\!=\!1.75$. At $\nu\!=\!41\pi/\beta\!\sim\!2.5$ the asymptotic functions (red lines) clearly deviate from the exact values (black lines) to which the former converge only for frequencies $\nu\!~\sim\!101\pi/\beta$. This is surprising since the former frequency ($\nu\!=\!41\pi/\beta$) corresponds to the choice of the inner box ($I_0$) with $N_{\text{inv}}\!=\!40$ for which both correction methods have already converged to the exact results in Figs.~\ref{fig:comparison_EXACTINV_M1_M2_Om0}-\ref{fig:comparison_gamma_U1}. Hence our asymptotic correction techniques unexpectedly provide excellent results even for very small intervals $I_0$ for which the structures of the vertex functions have not yet decayed to their respective asymptotic values.

\section{Conclusions}
\label{sec:conclusions}

We have presented two different methods for improving the numerical treatment of the Bethe-Salpeter equations within a DMFT (as well an AIM) calculation and, in particular, for extracting the local irreducible vertex $\Gamma_r^{\nu\nu'\omega}$. The latter can be obtained via a matrix inversion of the generalized impurity susceptibilities $\chi_r^{\nu\nu'\omega}$ w.r.t. the fermionic Matsubara frequencies $\nu$ and $\nu'$. In practice, however, the \textemdash in principle infinite \textemdash range of these frequencies has to be restricted to a finite interval $I_0$ of size $N_{\text{inv}}$ in which $\chi_r^{\nu\nu'\omega}$ is calculated "exactly". Since the numerical effort for this task rapidly grows with the matrix size, the calculations are typically restricted to a relatively small $I_0$: this introduces an error in the determination of $\Gamma_r^{\nu\nu'\omega}$, which also affects the low frequency sector ($|\nu |$,  $|\nu' | < N_{\text{inv}}/2$).

The two methods described in this work, which represent a significant extension of previous approaches\cite{Kunes2011}, aim to mitigate \textemdash or even to completely remove \textemdash these inversion errors by downfolding the high-frequency contributions of $\chi_r^{\nu\nu'\omega}$ and $\Gamma_r^{\nu\nu'\omega}$ into the low-frequency interval $I_0$, and are formulated for being applicable to any scattering channel ($r$) and any value of the transfer bosonic frequency ($\omega$).
Both procedures lead to (additive) correction terms w.r.t. the plane inversion on $I_0$, which substantially improve the final result for $\Gamma_r^{\nu\nu'\omega}$. The important point exploited in these procedures is that the high-frequency functions $\Gamma_{r,\text{asym}}^{\nu\nu'\omega}$ and $\chi_{r,\text{asym}}^{\nu\nu'\omega}$ can be obtained numerically at a much lower cost (i.e., growing only linearly with the number of frequencies) compared to the respective exact expressions. In fact, they are defined by the physical response functions $\chi_r^{\omega}$ and the fermion-boson vertices $\lambda_r^{\nu\omega}$, which depend on {\sl only one} bosonic and (at most) only one fermionic  Matsubara frequency.

Interestingly, the correction terms provided by both methods are essentially equivalent in size, allowing the users to freely choose the one which better matches their DMFT/AIM algorithm of choice. Furthermore, the computed corrections are almost independent of the fermionic frequencies $\nu$ and $\nu'$ as well as of the bosonic frequency $\omega$. In fact, the error introduced by inverting $\chi_r^{\nu\nu'\omega}$ in a finite frequency range corresponds \textemdash to a large extent \textemdash to a rigid shift of the vertex function, which is properly compensated by using our newly introduced methods. Remarkably, the correction terms provide very accurate results even if the inner interval is restricted to frequencies where the two-particle correlation functions have not fully reached their asymptotic values.

As a testbed example, we have applied our method to the DMFT solution of a single band Hubbard model in parameter regimes, where numerically reliable results for $\Gamma_r^{\nu\nu'\omega}$ are available for comparison. The full strength of these methods will, however, become evident by considering more challenging situations, such as, e.g., calculations (i) in the low-$T$ regime (where more Matsubara frequencies are needed) and (ii) of multi-orbital systems (where the number of frequencies, for which $\chi_r^{\nu\nu'\omega}$ can be directly computed, is substantially limited). Moreover, our analysis allows for a generalization to treat the case of fully momentum-dependent BS equations for all systems with an instantaneous microscopic interaction. In this respect, our newly developed techniques will represent a valuable tool both for computing physical response functions at the DMFT level and for including non-local correlations effects by means of diagrammatic extensions of DMFT.

\acknowledgments
The authors thank A. Antipov, E. Gull, C. Hille, P. Hansmann, K. Held, T. Sch\"afer, for valuable discussions, and the Wolfgang Pauli Institute for the kind hospitality. We acknowledge financial support from the Russian Science Foundation through
Grant No. 16-42-01057, the Deutsche Forschungsgemeinschaft
(DFG) through ZUK 63 and Projects No. AN 815/5-1 and No. AN 815/6-1, and the Austrian Science Fund (FWF) within the Project F41 (SFB ViCoM)
and the Project I 2794-N35. Calculations were performed on the
Vienna Scientific Cluster (VSC).

\appendix

\onecolumngrid
\section{Derivation of the vertex asymptotics}
\label{app:derivasympt}

\begin{figure}[t]
	\centering
		\includegraphics[width=0.7\textwidth]{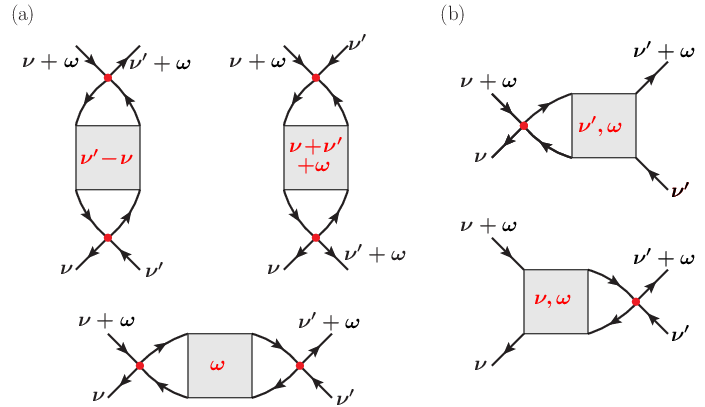}
	\caption{Diagrams contributing to the full vertex $F_r^{\nu\nu'\omega}$. Red dots denote the bare Hubbard interaction $U$. The characteristic frequency combination on which the diagram depends is marked in red. If the gray box represents the full vertex $F_r^{\nu\nu'\omega}$ the diagrams in (a) correspond to the physical susceptibility [Eq.~(\ref{equ:susphysicdm})], while the ones in (b) are related to the fermion-boson vertex [Eq.~(\ref{equ:fermbosvert})].}
	\label{fig:diagrams}
\end{figure}

The diagrammatic techniques to analyze the high-frequency behavior of the vertex functions $F_r^{\nu\nu'\omega}$ and $\Gamma_r^{\nu\nu'\omega}$ have been developed and discussed extensively in Refs.~[\onlinecite{Rohringer2012,Hummel2014,Wentzell2016a}]. In the following, we will only recall the basic concepts which are relevant for the present work. The key to the high-frequency behavior of the vertices relies on the following observation: If an external particle (hole) with an energy $\nu$ enters a (Feynman) diagram for the two-particle scattering amplitude at a vertex $U$, which is otherwise connected only to (three) internal propagators, the frequency $\nu$ will appear in one or more of the internal Green's functions. Consequently, the $\nu$-dependence of such a diagram will follow the $\nu$-dependence of these Green's functions and, hence, it decays for $\nu\!\rightarrow\!\infty$ at least as $1/i\nu$. The situation is different, if two external particles (or a particle and a hole) with frequencies $\nu_1$ and $\nu_2$ are scattered at the {\sl same} bare vertex $U$: In this case, the diagram will depend only on the combination $\nu_1\!\pm\!\nu_2$. As a result, its contribution to the vertex functions will remain finite even when $\nu_1,\nu_2\!\rightarrow\!\infty$, as long as $\nu_1\!\pm\!\nu_2$ is kept fixed. This observation allows to classify\cite{Wentzell2016a,Rohringer2018} all Feynman diagrams for $F_r^{\nu\nu'\omega}$ and $\Gamma_r^{\nu\nu'\omega}$ in the following way: (i) The first class consists of all diagrams where {\sl both} the two incoming and the two outgoing particles (holes) enter at the respective same bare vertices $U$, see Fig.~\ref{fig:diagrams}(a). According to the considerations above, these diagrams, hence, depend only on a {\sl single} (bosonic) combination of the incoming and outgoing frequencies, rather than on each of them independently. From Fig.~\ref{fig:diagrams}(a) one can see, that the contribution of all diagrams of this type obviously corresponds to the physical susceptibilities defined in Eqs.~(\ref{equ:susphysicdm}). (ii) For the second class of diagrams either the incoming {\sl or} the outgoing lines enter at the same bare vertex, see Fig.~\ref{fig:diagrams}(b). These diagrams depend only on one bosonic and one fermionic Matsubara frequency and can be related to a fermion-boson vertex\cite{Ayral2016a,Rohringer2016,Wentzell2016a,Rohringer2018} defined in Eqs.~(\ref{equ:fermbosvert}). (iii) For the third class of diagrams, all four external particles enter at different bare vertices and for this reason the corresponding diagrams decay at large frequencies and do not contribute to the asymptotics.

The crucial point is now that all diagrams of class (i) and (ii) \textemdash which are responsible for the non-trivial high-frequency asymptotic behavior of the vertex \textemdash are reducible in one of the three channels ($ph$, $\overline{ph}$, $pp$). In fact, for two outer lines entering the same bare vertex $U$ only two inner lines can be attached to it. As a consequence, when these two inner lines are cut the corresponding bare vertex gets separated from the rest of the diagram which is, hence, (two-particle) reducible. Let us denote all diagrams reducible in a given channel $r$ by the vertex function $\Phi_r^{\nu\nu'\omega}$, with $r\!=\!d,m,s,t$ (assuming the natural frequency notation, i.e., $ph$ for $r\!=\!d,m$ and $pp$ for $r\!=\!s,t,$). Obviously, the sum of all reducible and irreducible diagrams in a given channel yields all two-particle diagrams, i.e., $\Gamma_r^{\nu\nu'\omega}\!+\!\Phi_r^{\nu\nu'\omega}\!=\!F_r^{\nu\nu'\omega}$. Moreover, since each diagram is either (fully) two-particle irreducible or reducible in exactly one channel\cite{Rohringer2012}, $\Gamma_r^{\nu\nu'\omega}$ corresponds to the sum of all diagrams which are either fully irreducible or reducible in a channel $r'\!\ne\!r$. From this follow the so-called parquet equations

\begin{subequations}
\label{eq:parquet_eqs}
\begin{align}
\label{equ:parquetchanneld}
\Gamma_d^{\nu\nu'\omega}&=\Lambda_d^{\nu\nu'\omega}-\frac{1}{2}\Phi_d^{\nu(\nu+\omega)(\nu'-\nu)}-\frac{3}{2}\Phi_m^{\nu(\nu+\omega)(\nu'-\nu)}+\frac{1}{2}\Phi_s^{\nu\nu'(\nu+\nu'+\omega)}+\frac{3}{2}\Phi_t^{\nu\nu'(\nu+\nu'+\omega)},\\
\label{equ:parquetchannelm}
\Gamma_m^{\nu \nu' \omega} &= \Lambda_{m}^{\nu \nu' \omega} - \frac{1}{2} \Phi_{d}^{\nu (\nu+\omega) (\nu'-\nu)} + \frac{1}{2} \Phi_{m}^{\nu (\nu+\omega)(\nu'-\nu)} - \frac{1}{2} \Phi_{s}^{\nu \nu' (\nu+\nu'+\omega)} + \frac{1}{2} \Phi_{t}^{\nu \nu' (\nu+\nu'+\omega)},\\
\label{equ:parquetchannels}
\Gamma_s^{\nu \nu' \omega} &= \Lambda_{s}^{\nu \nu' \omega} + \frac{1}{2} \Phi_{d}^{\nu (\omega-\nu') (\nu'-\nu)} - \frac{3}{2} \Phi_{m}^{\nu (\omega-\nu')(\nu'-\nu)} + \frac{1}{2} \Phi_{d}^{\nu \nu' (\omega - \nu -\nu')} - \frac{3}{2} \Phi_{m}^{\nu \nu' (\omega-\nu-\nu')},\\
\label{equ:parquetchannelt}
\Gamma_t^{\nu \nu' \omega} &= \Lambda_{t}^{\nu \nu' \omega} - \frac{1}{2} \Phi_{d}^{\nu (\omega-\nu') (\nu'-\nu)} - \frac{1}{2} \Phi_{m}^{\nu (\omega-\nu')(\nu'-\nu)} + \frac{1}{2} \Phi_{d}^{\nu \nu' (\omega - \nu -\nu')} + \frac{1}{2} \Phi_{m}^{\nu \nu' (\omega-\nu-\nu')},
\end{align}
\end{subequations}

where $\Lambda_r^{\nu\nu'\omega}$ denotes the fully irreducible vertex which does not exhibit any dependence on the (irreducibility) channels $ph$, $\overline{ph}$ and $pp$. Hence, here the index $r$ labels \textemdash as for $F_r^{\nu\nu'\omega} $ and $\chi_r^{\nu\nu'\omega}$\textemdash only the spin combination and the frequency convention ($ph$ for $r\!=\!d,m$ and $pp$ for $r\!=\!s,t$) in which $\Lambda_r^{\nu\nu'\omega}$ is represented. Note that on a first glance it seems rather inconsistent that the vertices reducible in the density and spin (i.e., in the $ph$) channels, $\Phi_d^{\nu\nu'\omega}$ and $\Phi_m^{\nu\nu'\omega}$, contribute to the vertices which are irreducible exactly in the same channels ($\Gamma_d^{\nu\nu'\omega}$ and $\Gamma_m^{\nu\nu'\omega}$). The reason for this contradiction is that these contributions originate from the transverse particle-hole channel ($\overline{ph}$) which, of course, contributes to $\Gamma_d^{\nu\nu'\omega}$ and $\Gamma_m^{\nu\nu'\omega}$. Considering crossing and SU(2) symmetry, $\Phi_{\overline{ph},\sigma\sigma'}^{\nu\nu'\omega}$ can be represented by the corresponding functions in the longitudinal ($ph$) channel by means of a frequency shift which explains the presence of $\Phi_d^{\nu(\nu+\omega)(\nu'-\nu)}$ and $\Phi_m^{\nu(\nu+\omega)(\nu'-\nu)}$ in Eqs.~(\ref{equ:parquetchanneld}) and (\ref{equ:parquetchannelm}).

According to the discussion above, the asymptotic high-frequency behavior of $\Gamma_r^{\nu\nu'\omega}$ is determined by the asymptotic behavior of the reducible vertices $\Phi_r$ in Eqs.~(\ref{eq:parquet_eqs}). The latter in turn is given by the diagrams in Fig.~\ref{fig:diagrams}(a) [for the $ph$ ($r\!=\!d,m$) channels by the two upper and for the $pp$ ($r\!=\!s,t$) channels by the upper left and the lower diagrams] which can be expressed in terms of the physical susceptibilities [Eq.~(\ref{equ:susphysicdm})] as

\begin{subequations}
\label{equ:K1asympt}
\begin{align}
\label{equ:K1asymptdens}
 &\Phi_{d,\text{asym}}^{\nu bc}= -U^2\chi_d^{c}\\
\label{equ:K1asymptmagn}
 &\Phi_{m,\text{asym}}^{\nu bc}= -U^2\chi_m^{c}\\
\label{equ:K1asymptsing}
&\Phi_{s,\text{asym}}^{\nu\nu'(\nu+\nu'+\omega)}=-2U^2\chi_{pp,\uparrow\downarrow}^{\nu+\nu'+\omega}\\
\label{equ:K1asympttrip}
&\Phi_{t,\text{asym}}^{\nu\nu'(\nu+\nu'+\omega)}=0,
\end{align}
\end{subequations}

where $b$ and $c$ represent the respective frequency arguments according to Eqs.~\ref{eq:parquet_eqs}. Note, that for the derivation of the asymptotic behavior of $\Phi_r$ in Eqs.~(\ref{equ:K1asympt}) the bosonic frequency $\omega$ (in the respective natural notation of the given channel) has been considered to be fixed. For this reason, no asymptotic contributions from fermion-boson diagrams [as given in Fig.~\ref{fig:diagrams}(b)] can arise since the latter would be constant along lines in the three-dimensional frequency space which are not parallel to planes of constant $\omega$. Inserting the high-frequency expressions for $\Phi_r$ in Eqs.~(\ref{equ:K1asympt}) into the parquet equations for $\Gamma_r$ [Eqs.~(\ref{eq:parquet_eqs})] yields the asymptotic high-frequency functions $\Gamma_{r,\text{asym}}^{\nu\nu'\omega}$ (for a fixed $\omega$) given in Eqs.~(\ref{equ:Gammahighfreq}) in Sec.~\ref{subsec:asympt}.

For the determination of $F_{r,\text{asym}}^{\nu\nu'\omega}$ we use the fact $F_r^{\nu\nu'\omega}=\Gamma_r^{\nu\nu'\omega}\!+\Phi_r^{\nu\nu'\omega}$. Hence, in addition to the high-frequency behavior of $\Gamma_r^{\nu\nu'\omega}$ discussed above [and explicitly given in Eq.~(\ref{equ:Gammahighfreq})] we have to determine $\Phi_{r,\text{asym}}^{\nu\nu'\omega}$. Let us stress that the latter is different from the expressions given in Eq.~(\ref{equ:K1asympt}) due to the difference in the frequency arguments. Specifically, for $\Phi_{r,\text{asym}}^{\nu\nu'\omega}$ (i.e., without any shifts in the arguments) for $\nu,\nu'\!\rightarrow\!\infty$ one has to consider contributions of diagrams such as given in Fig.~\ref{fig:diagrams}(b) which correspond to fermion-boson vertices [Eq.~(\ref{equ:fermbosvert})]. Explicitly, we obtain

\begin{subequations}
\label{eq:phi_asy}
\begin{align}
 \label{eq:phi_asyd}
 &\Phi_{d,\text{asym}}^{\nu\nu'\omega}=U\lambda_d^{\nu\omega}+U\lambda_d^{\nu'\omega}+U^2\chi_d^{\omega},\\
 \label{eq:phi_asym}
 &\Phi_{m,\text{asym}}^{\nu\nu'\omega}=U\lambda_m^{\nu\omega}+U\lambda_m^{\nu'\omega}+U^2\chi_{m}^{\omega},\\
 \label{eq:phi_asys}
 &\Phi_{s,\text{asym}}^{\nu\nu'\omega}=2U\lambda_{pp,\uparrow\downarrow}^{\nu\omega}+2U\lambda_{pp,\uparrow\downarrow}^{\nu'\omega}+2U^2\chi_{pp,\uparrow \downarrow}^{\omega},\\
 \label{eq:phi_asyt}
 &\Phi_{t,\text{asym}}^{\nu\nu'\omega}=0,
\end{align}
\end{subequations}

where the terms $\chi_r^{\omega}$ remove the double counting of contributions which are contained in both $\lambda_r^{\nu\omega}$ and $\lambda_r^{\nu'\omega}$ [cf. the two diagrams in Fig.~\ref{fig:diagrams} which both contain the lower diagram in Fig.~\ref{fig:diagrams}(a)]. Adding the asymptotic contributions for $\Phi_{r,\text{asym}}^{\nu\nu'\omega}$ in Eqs.~(\ref{eq:phi_asy}) to the corresponding ones of $\Gamma_{r,\text{asym}}^{\nu\nu'\omega}$ in Eq.~(\ref{equ:Gammahighfreq}) yields $F_{r,\text{asym}}^{\nu\nu'\omega}$ as given in Eq.~(\ref{eq:f_eqs_asy}).


\section{Numerical calculation of $\chi_r^\omega$ and $\lambda_r^{\nu\omega}$ in ED}
\label{app:calckernelfunc}

As mentioned in Section \ref{sec:theory}, the asymptotic functions of the two-particle vertex, i.e., the susceptibility and the fermiion-boson vertex, can be evaluated in the same way as the two-particle Green's function by means of the impurity solver used in our (ED)DMFT cycle. In fact, while the treatment of the full frequency dependence of the vertex is computationally challenging, valuable information of the two-particle scattering processes can be extracted from more manageable quantities whose parameter dependence is restricted to one or two frequencies [in our single-band SU(2) symmetric Hubbard model]. In this section, we explicitly derive their spectral representation, as implemented in our impurity solver.

\subsubsection{Lehmann representation of $\lambda_r^{\nu\omega}$}

In this appendix, we derive the Lehmann representation for the fermion-boson vertices $\lambda_r^{\nu'\omega}$. {\sl First}, we present the equations for the $ph$ channel, i.e., for $\lambda_{d/m}^{\nu'\omega}\!=\!\lambda_{ph,\uparrow\uparrow}^{\nu'\omega}\!\pm\!\lambda_{ph,\uparrow\downarrow}^{\nu\omega}$. Our starting point is the Fourier representation of the generalized susceptibility\cite{Toschi2007,Hafermann2009} $\chi_r^{\nu\nu'\omega}$ summed over one fermionic Matsubara frequency ($\nu$)

\begin{equation}
\begin{split}
\label{eq:lambdadef}
\tilde{\lambda}_{ph, \sigma \sigma'}^{\nu' \omega}=\frac{1}{\beta}\sum_{\nu}\chi_{ph,\sigma\sigma'}^{\nu\nu'\omega}=\frac{1}{\beta}\sum_{\nu} \int\limits_{0}^{\beta}&d\tau_{1}d\tau_{2}d\tau_{3}\, e^{-i\nu\tau_{1}}e^{i(\nu+\omega)\tau_{2}}e^{-i(\nu'+\omega)\tau_{3}} \\ \times&\big[\big<T_{\tau}c_{\sigma}^{\dagger}(\tau_{1})c_{\sigma}(\tau_{2})c_{\sigma'}^{\dagger}(\tau_{3})c_{\sigma'}(0)\big>-\big<T_{\tau}c_{\sigma}^{\dagger}(\tau_{1})c_{\sigma}(\tau_{2})\big>\big<T_{\tau}c_{\sigma'}^{\dagger}(\tau_{3})c_{\sigma'}(0)\big>\big],
\end{split}
\end{equation}

which is related to the fermion-boson vertices $\lambda_{ph,\sigma\sigma'}^{\nu'\omega}$ as (note the inversion of $\sigma'$)
\begin{equation}
\lambda_{ph,\sigma\sigma'}^{\nu'\omega}=-\frac{\tilde{\lambda}_{ph,\sigma(-\sigma')}^{\nu'\omega}}{G(\nu')G(\nu'+\omega)}-\delta_{\sigma(-\sigma')}.
\end{equation}
Exchanging frequency summation and (imaginary) time integration in Eq.~(\ref{eq:lambdadef}), we obtain [$\frac{1}{\beta}\sum_\nu e^{-i\nu\tau}\!=\!\delta(\tau)$]

\begin{equation}
\label{eq:lambdadef_step1}
\tilde{\lambda}_{ph, \sigma \sigma'}^{\nu' \omega}=\int_{0}^{\beta} d\tau_1 d\tau_3\;e^{i\omega \tau_1} e^{-i(\nu'+\omega)\tau_3}\big[\big<T_{\tau}c_{\sigma}^{\dagger}(\tau_{1})c_{\sigma}(\tau_{1})c_{\sigma'}^{\dagger}(\tau_{3})c_{\sigma'}(0)\big>-\big<T_{\tau}c_{\sigma}^{\dagger}(\tau_{1})c_{\sigma}(\tau_{1})\big>\big<T_{\tau}c_{\sigma'}^{\dagger}(\tau_{3})c_{\sigma'}(0)\big>\big].
\end{equation}

Let us begin by calculating first term in the square brackets on the r.h.s. of Eq.~(\ref{eq:lambdadef_step1}). Considering the time-ordering operator, we obtain the following two contributions:

\begin{equation}
\begin{split}
\label{eq:lambda_termA_step1}
\tilde{\lambda}_{ph, \sigma \sigma'}^{\nu' \omega}=\int_{0}^{\beta} d\tau_1 \Biggl[ &\int_{0}^{\tau_1} d\tau_3 e^{i\omega \tau_1} e^{-i(\nu' + \omega)\tau_3} \langle c_{\sigma}^{\dagger}(\tau_{1})c_{\sigma}(\tau_{1})c_{\sigma'}^{\dagger}(\tau_{3})c_{\sigma'}(0) \rangle+\\&\int_{\tau_1}^{\beta} d\tau_3 e^{i\omega \tau_1} e^{-i(\nu'+\omega)\tau_3} \langle c_{\sigma'}^{\dagger}(\tau_{3})c^{\dagger}_{\sigma}(\tau_1)c_{\sigma}(\tau_{1})c_{\sigma'}(0) \rangle \Biggr]
= \tilde{\lambda}_{13,\sigma \sigma'}^{\nu' \omega} + \tilde{\lambda}_{31, \sigma \sigma'}^{\nu' \omega}.
\end{split}
\end{equation}

The first term can be transformed by inserting a complete basis of the Hilbert space ($\mathds{1}=\sum_i\lvert i\rangle\langle i\rvert$) after each operator in the trace

\begin{equation}
\label{eq:lambda_term13}
\begin{split}
\tilde{\lambda}_{13,\sigma \sigma'}^{\nu' \omega} &= \int_{0}^{\beta} d\tau_1 \int_{0}^{\tau_1} d\tau_3  e^{i\omega \tau_1} e^{-i(\nu'+\omega)\tau_3} \langle n_{\sigma}(\tau_1) c^{\dagger}_{\sigma'} c_{\sigma'}(0) \rangle \\
&  = \frac{1}{Z} \int_{0}^{\beta} d\tau_1 \int_{0}^{\tau_1} d\tau_3 e^{i\omega \tau_1} e^{-i(\omega+\nu')\tau_3} \sum_{i,j,k} \langle i \vert e^{-\beta H} e^{H\tau_1} n_{\sigma} e^{-H\tau_1} \vert j \rangle \langle j \vert e^{H\tau_3} c^{\dagger}_{\sigma'} e^{-H\tau_3} \vert k \rangle \langle k \vert c_{\sigma'} \vert i \rangle  \\
&=  \frac{1}{Z} \sum_{i,j,k} \int_{0}^{\beta} d\tau_1 \int_{0}^{\tau_1} d\tau_3 e^{i\omega \tau_1} e^{-i(\omega +\nu')\tau_3} e^{-\beta E_i} e^{E_i \tau_1} e^{-E_j \tau_1} e^{E_j \tau_3} e^{-E_k \tau_3} \langle i \vert n_{\sigma} \vert j \rangle \langle j \vert c^{\dagger}_{\sigma'} \vert k \rangle \langle k \vert c_{\sigma'} \vert i \rangle \\
& = \frac{1}{Z} \sum_{i,j,k} \frac{\langle i \vert n_{\sigma} \vert j \rangle \langle j \vert c^{\dagger}_{\sigma'} \vert k \rangle \langle k \vert c_{\sigma'} \vert i \rangle}{i(\nu'+\omega)+E_k-E_j} \Biggl[\frac{e^{-E_j \beta} - e^{-E_i \beta}}{i\omega + E_i - E_j} - \frac{e^{-E_k \beta} + e^{-E_i \beta}}{i\nu' + E_k - E_i } \Biggr],
\end{split}
\end{equation}

where $Z$ denotes the partition function (see Sec.~\ref{sec:theory}). Using the same procedure for $\tilde{\lambda}_{31,\sigma \sigma'}^{\nu' \omega}$ one obtains

\begin{equation}
\label{lambda_term31}
\begin{split}
\tilde{\lambda}_{31,\sigma \sigma'}^{\nu' \omega} =  \frac{1}{Z} \beta \sum_{i,j,k} \frac{\langle i \vert c^{\dagger}_{\sigma'} \vert j \rangle \langle j \vert n_{\sigma} \vert k \rangle \langle k \vert c_{\sigma'} \vert i \rangle}{i(\nu'+\omega)+E_j-E_i} \Biggl[\frac{e^{-E_k \beta} - e^{-E_j \beta}}{i\omega + E_j - E_k} + \frac{e^{-E_k \beta} + e^{-E_i \beta}}{i\nu' + E_k - E_i } \Biggr].
\end{split}
\end{equation}

Let us now consider the second term in the square brackets on the r.h.s. of Eq. \eqref{eq:lambdadef_step1}. It contains only one-particle correlation functions and can be rewritten as

\begin{equation}
\label{eq:lambda_termB}
\begin{split}
\tilde{\lambda}_{0,ph,\sigma \sigma'}^{\nu' \omega}=\int_{0}^{\beta} d\tau_1 \int_{0}^{\beta} d\tau_3 e^{i\omega \tau_1} e^{-i(\nu'+\omega)\tau_3} \langle T_{\tau} c^{\dagger}_{\sigma}(\tau_1) c_{\sigma}(\tau_1) \rangle \langle T_{\tau} c^{\dagger}_{\sigma'}(\tau_3) c_{\sigma'}(0) \rangle=\delta_{\omega 0} \langle n_{\sigma} \rangle G_{\sigma'}(\nu'),
\end{split}
\end{equation}

where $\langle n_{\sigma} \rangle$ is the local electron density which, at half filling and in the SU(2) symmetric case, simplifies to $ \langle n_{\sigma}\rangle\!=\!1/2$.

In the {\sl second} step, we present the derivation of $\lambda_{pp,\sigma\sigma'}^{\nu'\omega}$. This can be easily achieved by using the frequency transformation between $ph$ and $pp$ channels. As in the $ph$ case we define a quantity

\begin{equation}
\label{eq:lambda_ppdef}
\begin{split}
\tilde{\lambda}_{pp, \sigma \sigma'}^{\nu' \omega} & =\frac{1}{\beta} \sum_{\nu}  \chi^{\nu\nu'\omega}_{{pp},\sigma\sigma'}\!\equiv\! \frac{1}{\beta}\sum_{\nu} \chi^{\nu\nu'(\omega-\nu-\nu')}_{{ph},\sigma\sigma'} \\
& = \int_{0}^{\beta} d\tau_1 d\tau_2 e^{i(\omega-\nu')\tau_2} e^{-i \omega\tau_1} \Bigl[ \langle T_{\tau} c^{\dagger}_{\sigma}(\tau_1)c_{\sigma}(\tau_2)c^{\dagger}_{\sigma'}(\tau_1)c_{\sigma'}(0) \rangle - \langle T_{\tau} c^{\dagger}_{\sigma}(\tau_1) c_{\sigma}(\tau_2) \rangle \langle T_{\tau} c^{\dagger}_{\sigma'}(\tau_1) c_{\sigma'}(0) \rangle \Bigr],
\end{split}
\end{equation}

which is related to $\lambda_{pp,\sigma\sigma'}^{\nu'\omega}$ via $\lambda_{pp,\sigma\sigma'}^{\nu'\omega}=\tilde{\lambda}_{pp,\sigma\sigma'}^{\nu'\omega}/[G(\nu')G(\omega-\nu')]$. In analogy to the previous derivation, one can express the first term of the second line of Eq.~(\ref{eq:lambda_ppdef}) as a sum of two contributions:

\begin{equation}
\label{eq:lambdapp_termA}
\tilde{\lambda}_{pp,\sigma \sigma'}^{\nu' \omega} = \tilde{\lambda}_{12, \sigma \sigma'}^{\nu' \omega} + \tilde{\lambda}_{21, \sigma \sigma'}^{\nu' \omega},
\end{equation}

with $\tau_1 > \tau_2 $ and $\tau_2 > \tau_1$, respectively (we omitted the $pp$ label on the r.h.s. to simplify the notation). The explicit expressions then read:

\begin{subequations}
\begin{align}
\label{eq:lambdapp_12}
\tilde{\lambda}_{12, \sigma \sigma'}^{\nu' \omega} &=  \frac{1}{Z}\beta \sum_{i,j,k} \frac{\langle i \vert \Delta_{\sigma \sigma'}^{*} \vert j \rangle \langle j \vert c_{\sigma} \vert k \rangle \langle k \vert c_{\sigma'} \vert i \rangle}{i(\omega -\nu')+ E_j - E_k} \Biggl[ \frac{e^{-E_i \beta} - e^{-E_j \beta}}{i\omega+E_j-E_i} - \frac{e^{-E_k \beta} + e^{-E_i \beta}}{i\nu' + E_k - E_i} \Biggr]\\
\tilde{\lambda}_{21, \sigma \sigma'}^{\nu' \omega} & = - \frac{1}{Z} \beta \sum_{i,j,k} \frac{\langle i c_{\sigma} \vert j \rangle \langle j \vert \Delta_{\sigma \sigma'}^{*} \vert k \rangle \langle k \vert c_{\sigma'} \vert i \rangle}{i(\omega -\nu')+ E_i - E_j} \Biggl[ \frac{e^{-E_k \beta} - e^{-E_j \beta}}{i\omega+E_k-E_j} - \frac{e^{-E_k \beta} + e^{-E_i \beta}}{i\nu' + E_k - E_i} \Biggr],
\end{align}
\end{subequations}

with $\Delta_{\sigma\sigma'}^{(*)} = c_{\sigma}^{(\dagger)} c_{\sigma'}^{(\dagger)}$ representing the pair annihilation (creation) operator. The second term of Eq. \eqref{eq:lambda_ppdef}, in the following referred as $\tilde{\lambda}_{0,pp,\sigma\sigma'}^{\nu' \omega}$, can be expressed by means of single-particle propagators:

\begin{equation}
\label{eq:lambdapp_termB}
\tilde{\lambda}_{0,pp,\sigma\sigma'}^{\nu' \omega}=\int_{0}^{\beta} d\tau_1 d\tau_2 e^{i(\omega -\nu')\tau_2} e^{-i\omega \tau_1} G_{\sigma}(\tau_1- \tau_2) G_{\sigma'}(\tau_1)=G(\nu')G(\omega-\nu').
\end{equation}

\subsubsection{Lehmann representation of $\chi_{r}^{\omega}$}

In this section we derive the spectral representation for the susceptibilities \eqref{equ:susphysicdm}. In particular, we consider its building blocks

\begin{equation}
\chi_{ph/pp, \sigma \sigma'}^{\omega} = \frac{1}{\beta^2}\sum_{\nu \nu'} \chi_{ph/pp, \sigma \sigma'}^{\nu \nu' \omega}.
\end{equation}

We start by analyzing the particle-hole channel:

\begin{equation}
\begin{split}
\label{eq:suscept_lehmann_ph}
\chi_{ph, \sigma \sigma'}^{\omega} &= \frac{1}{\beta^2}\sum_{\nu \nu'} \int_{0}^{\beta} d\tau_1 d\tau_2d\tau_3 e^{-i\nu \tau_1} e^{i(\nu + \omega) \tau_2} e^{-i(\nu'+\omega)\tau_3} \Bigl[ \langle T_{\tau} c^{\dagger}_{\sigma}(\tau_1) c_{\sigma}(\tau_2) c^{\dagger}_{\sigma'}(\tau_3) c_{\sigma'}(0) \rangle - \langle T_{\tau} c^{\dagger}_{\sigma}(\tau_1) c_{\sigma} (\tau_2) \rangle \langle T_{\tau} c^{\dagger}_{\sigma'}(\tau_3) c_{\sigma'}(0) \rangle \Bigr ]\\
&=   \int_{0}^{\beta} d\tau e^{i\omega \tau} \Bigl[ \langle T_{\tau} c^{\dagger}_{\sigma}(\tau) c_{\sigma}(\tau) c^{\dagger}_{\sigma'}(0) c_{\sigma'}(0) \rangle - \langle n_{\sigma}(\tau) \rangle \langle n_{\sigma'}(0) \rangle \Bigr],
\end{split}
\end{equation}

The actual derivation of the spectral representation is now completely analogous to the one for the fermion-boson vertex illustrated above and we obtain

\begin{equation}
\label{eq:suscept_ph_termA}
\chi_{ph, \sigma \sigma'}^{\omega} = \frac{1}{Z} \sum_{i,j} \frac{\langle i \vert n_{\sigma} \vert j \rangle \langle j \vert n_{\sigma'} \vert i \rangle}{i\omega +E_i -E_j} \Bigl(e^{-E_j \beta} - e^{-E_i \beta}\Bigr)-\beta\langle n_\sigma\rangle^2\delta_{\omega 0},
\end{equation}

where we have used that in the SU(2) symmetric case $\langle n_\uparrow\rangle\!=\!\langle n_\downarrow\rangle$. For the particle-particle channel one finds:

\begin{equation}
\label{eq:susept_pp_termA}
\chi_{pp, \sigma \sigma'}^{\omega} = \frac{1}{Z}\sum_{i,j} \frac{\langle i \vert \Delta_{\sigma \sigma'}^{*} \vert j \rangle \langle j \vert \Delta_{\sigma \sigma'} \vert i \rangle}{i\omega +E_j - E_i} \Bigl(e^{-E_i \beta} - e^{-E_j \beta}\Bigr).
\end{equation}

Let us finally emphasize the simplicity of the above expressions with respect to the Lehmann representation of the full (not summed) three-frequency two-particle propagator. The equal time evaluation significantly reduces the number of permutations due to the time-ordering. In addition, the reduced variable dependence cuts down the huge computational cost of evaluating the two-particle propagators.

\section{Self-consistent evaluation of $\chi_r^\omega$ and $\lambda_r^{\nu\omega}$ }
\label{app:calckernelfunc_selfcons}

\begin{figure}[b!]
\begin{center}
\subfloat{\includegraphics[width=0.65\textwidth]{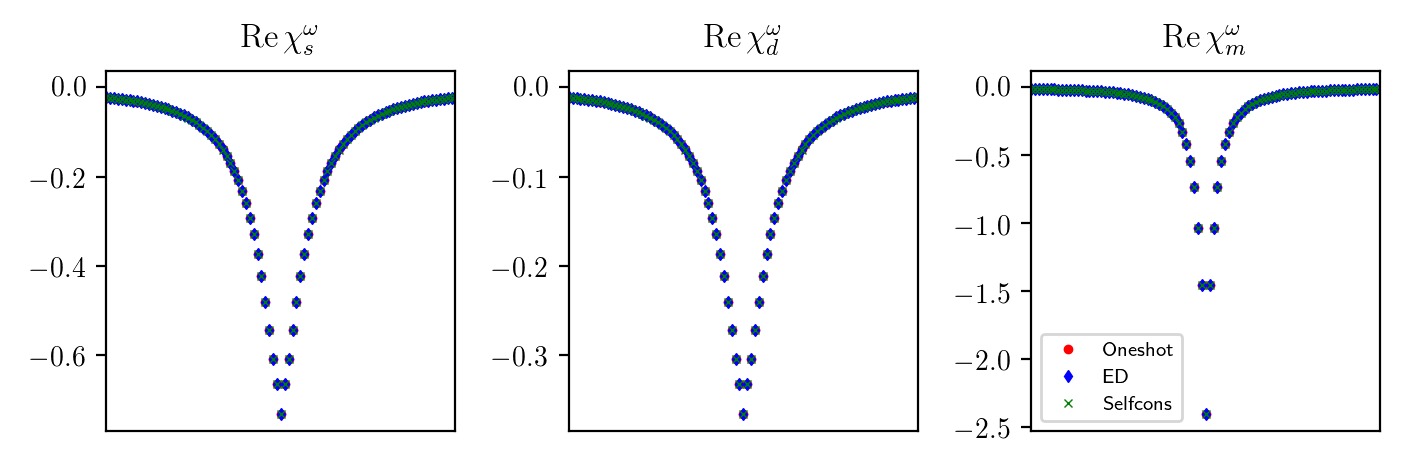}}\\
\subfloat{\includegraphics[width=0.67\textwidth]{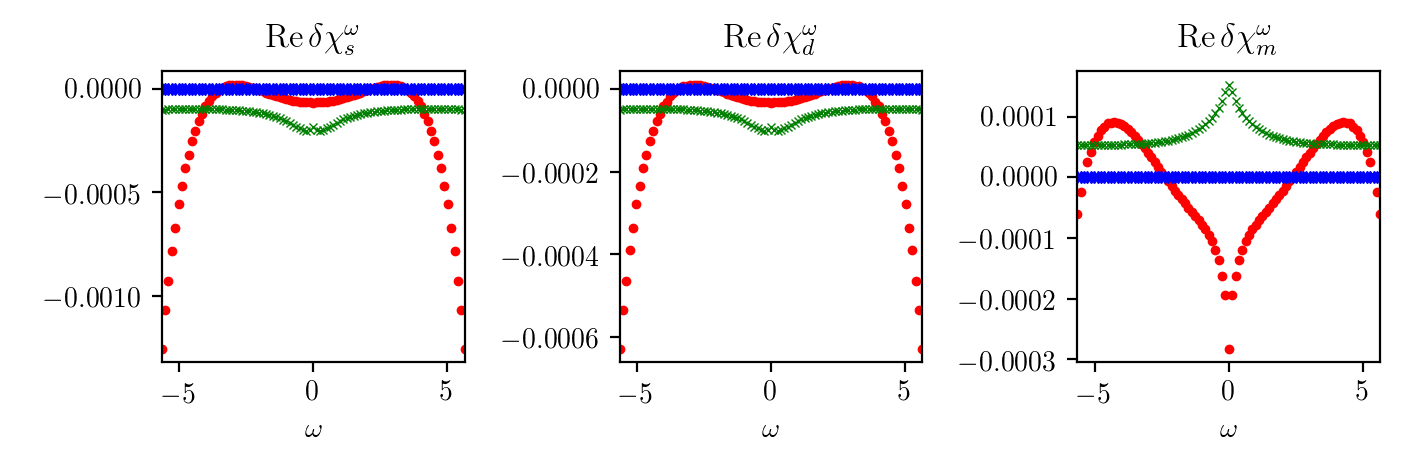}}
\caption{[Color online] Comparison of the physical susceptibilities acquired by means of the ED impurity solver (blue diamonds), by means of the self-consistent procedure (green crosses) and by directly summing the generalized susceptibilities $\chi_{r}^{\nu \nu' \omega}$ over $\nu$ and $\nu'$ (red dots). The first row shows the susceptibilities in the physical channels for $U=1$ while the second one displays the correspondent difference to the exact results.}
\label{fig:self_cons_asympt}
\end{center}
\end{figure}

\begin{figure}[b!]
\begin{center}
\subfloat{\includegraphics[width=0.65\textwidth]{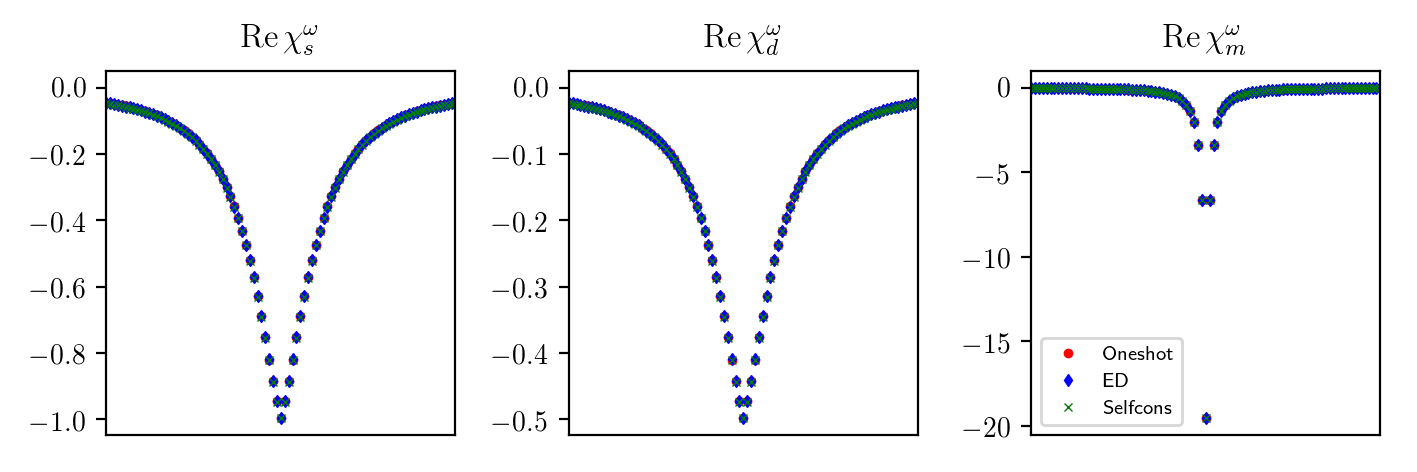}}\\
\subfloat{\includegraphics[width=0.67\textwidth]{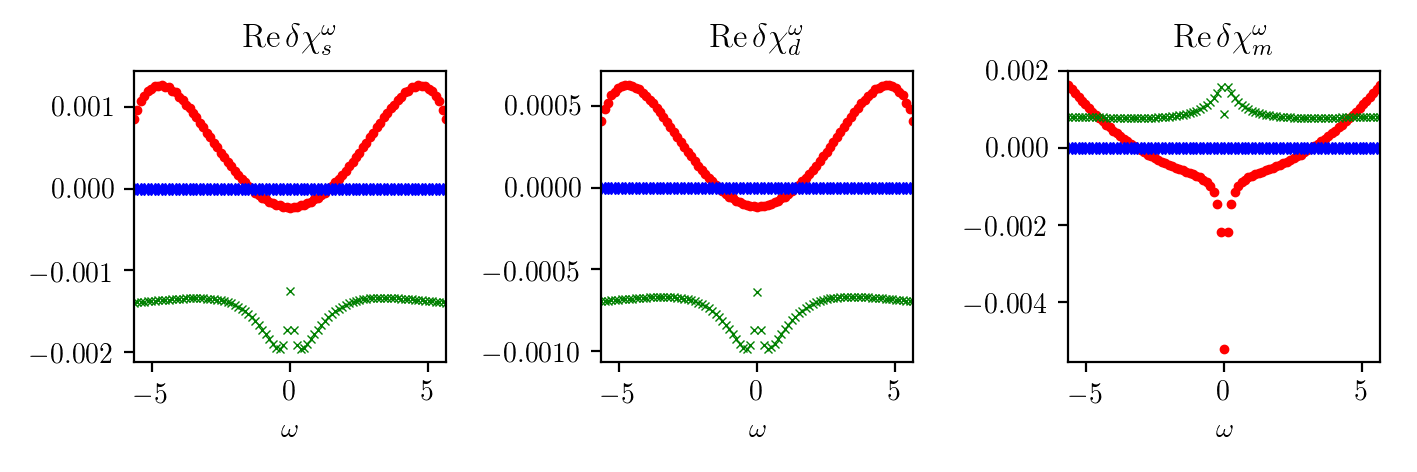}}
\caption{[Color online] Same as in Fig.~\ref{fig:self_cons_asympt} for $U=1.75$.}
\label{fig:self_cons_asympt_1p75}
\end{center}
\end{figure}

\begin{figure}[b!]
\begin{center}
\subfloat{\includegraphics[width=0.66\textwidth]{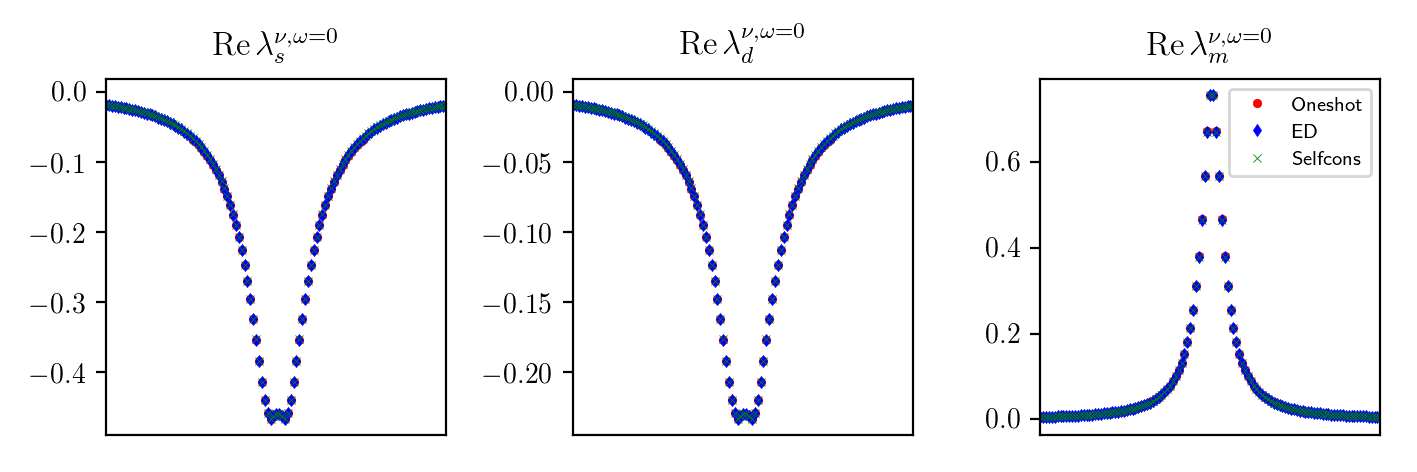}}\\
\subfloat{\includegraphics[width=0.66\textwidth]{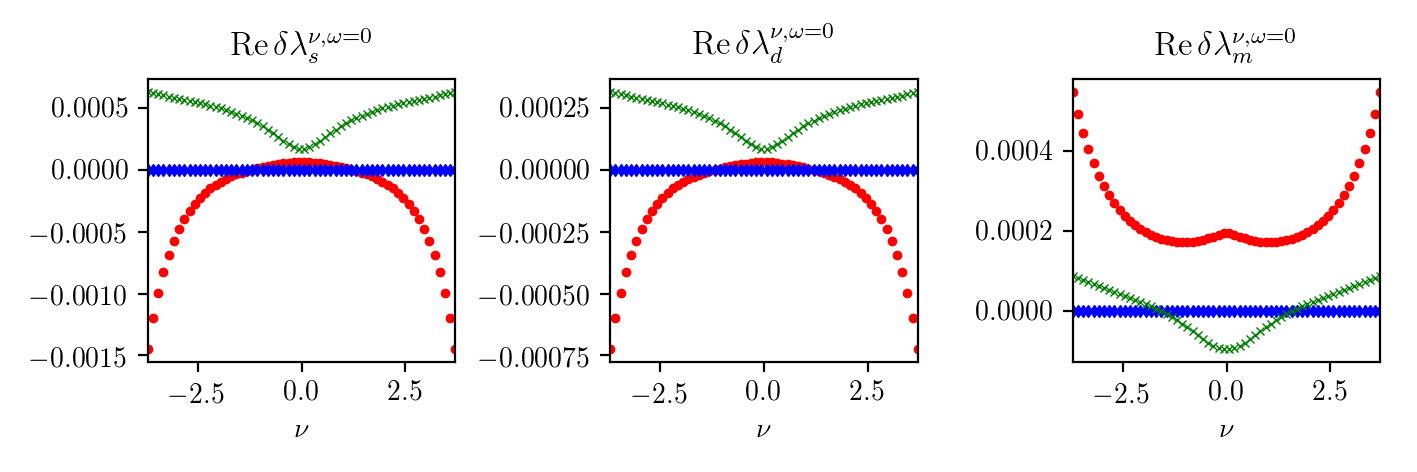}}
\caption{[Color online] Comparison of the fermi-boson vertices acquired by means of the ED impurity solver (blue diamonds), by means of the self-consistent procedure (green crosses) and by directly summing the generalized susceptibilities $\chi_{r}^{\nu \nu' \omega}$ over $\nu'$ (red dots). Here $\lambda^{\nu \omega}$ is plotted as a function of the fermionic frequency $\nu$ and for a fixed bosonic frequency $\omega = 0$. The first row shows $\lambda^{\nu\omega=0}$ in the physical channels for $U=1$ while the second one displays the corresponding difference to the exact results.}
\label{fig:self_cons_asympt_P}
\end{center}
\end{figure}

\begin{figure}[b!]
\begin{center}
\subfloat{\includegraphics[width=0.66\textwidth]{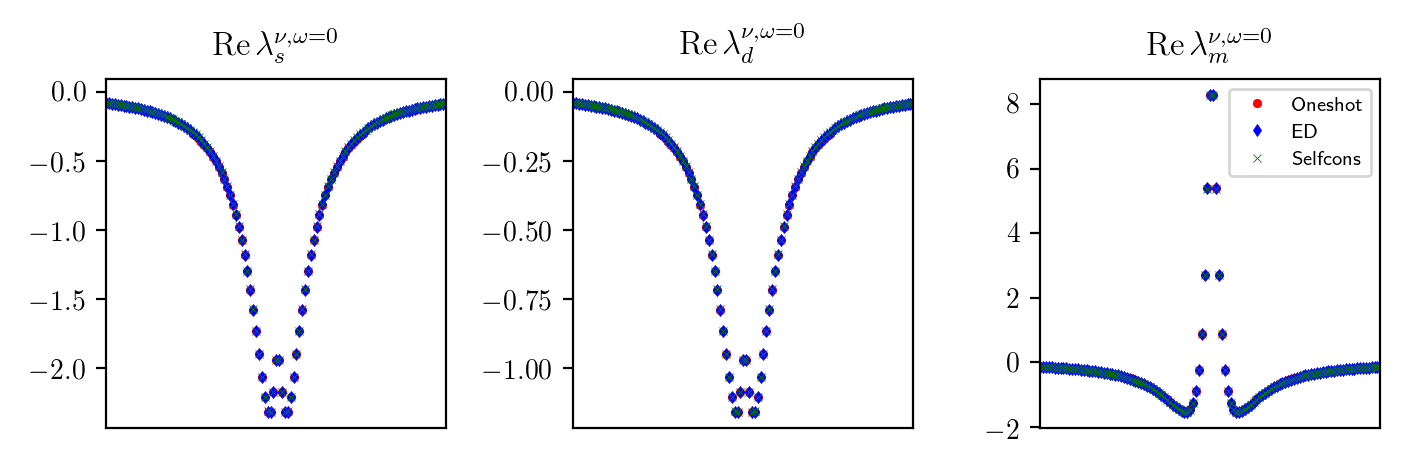}}\\
\subfloat{\includegraphics[width=0.66\textwidth]{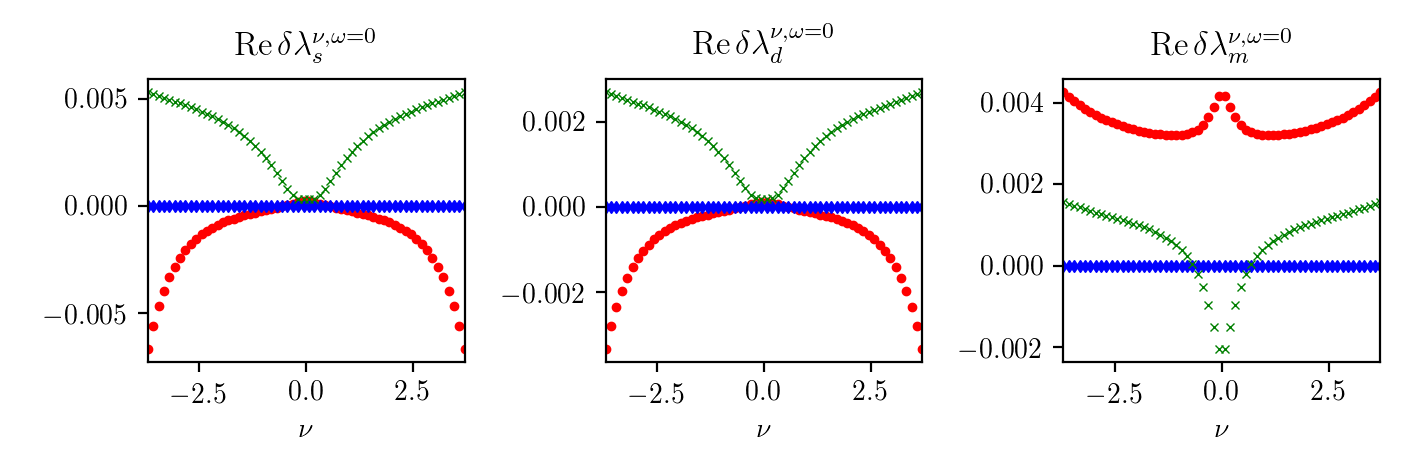}}
\caption{[Color online] Same as in Fig.~\ref{fig:self_cons_asympt_P} for $U=1.75$.}
\label{fig:self_cons_asympt_P_1p75}
\end{center}
\end{figure}

We remark, that although the spectral representation of $\chi_r^{\omega}$  and $\lambda_r^{\nu \omega}$ represents a computationally straightforward task for the ED impurity solver, alternative schemes to acquire the asymptotics of the vertex function are available. Besides the implementations used by other impurity solvers as Quantum Monte Carlo (QMC)\cite{Kaufmann2017}, we recall here another approach which turns out to be feasible for different vertex-based solvers as the D$\Gamma$A \cite{Toschi2007,Rohringer2016,Valli2015}, the parquet approximation \cite{Chen1992,Bickers1991a,Tam2013}, and the functional renormalization group (fRG) \cite{ Metzner2012,Taranto2014,Wentzell2016a,Halboth2000}. This technique, which has been proposed in Ref.~\onlinecite{Wentzell2016a}, is based on a self-consistent determination of the asymptotic functions from the low-frequency data for the two-particle Green's function (see, in particular, Appendix~C in Ref.\onlinecite{Wentzell2016a}). In Figs.~\ref{fig:self_cons_asympt}-\ref{fig:self_cons_asympt_P_1p75}, we show converged results which exhibit deviations to the exact value at most of the order of $\delta \sim 10^{-3}$ (relative error). Nevertheless, one should note that this approach presents some intrinsic drawbacks which may become pathological in certain parameter regimes, in particular for too small frequency ranges which do not capture the entire low-frequency structures of the two-particle correlation functions.

\section{Comparison of the two methods for U=1.75}
\label{app:gamma_1.75}

\begin{figure}[t!]
\begin{center}
\subfloat{\includegraphics[width=0.5\textwidth]{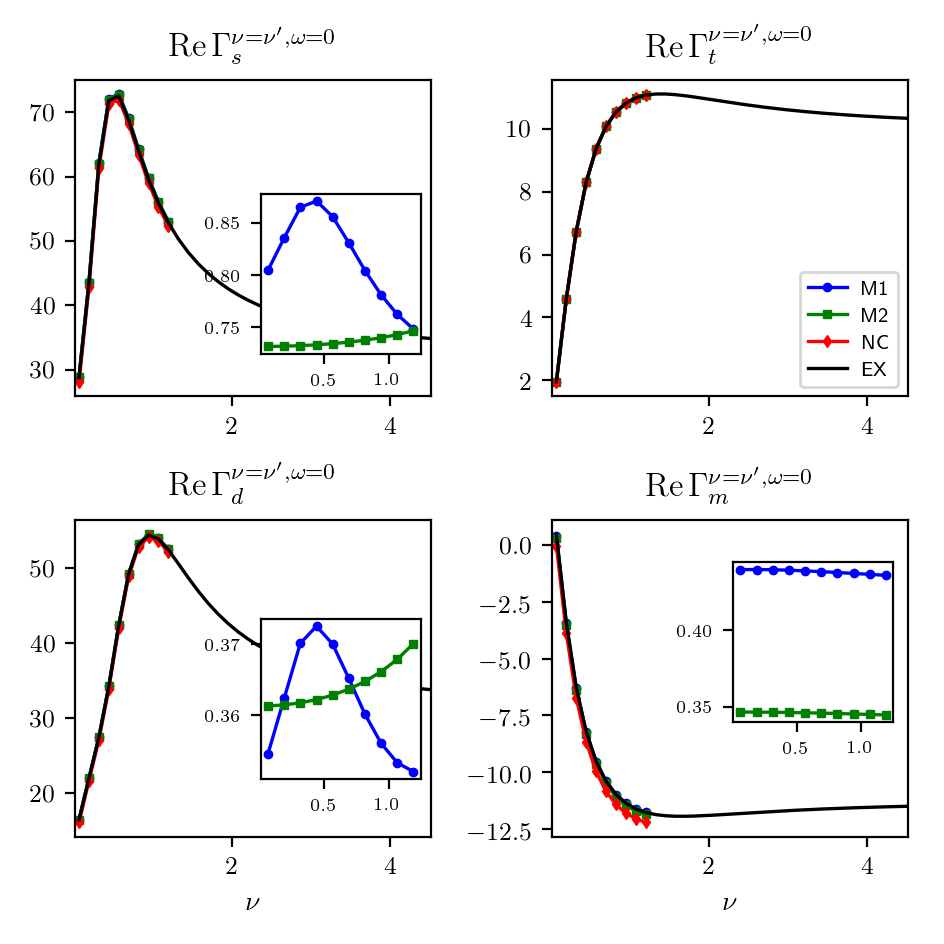}}
\subfloat{\includegraphics[width=0.5\textwidth]{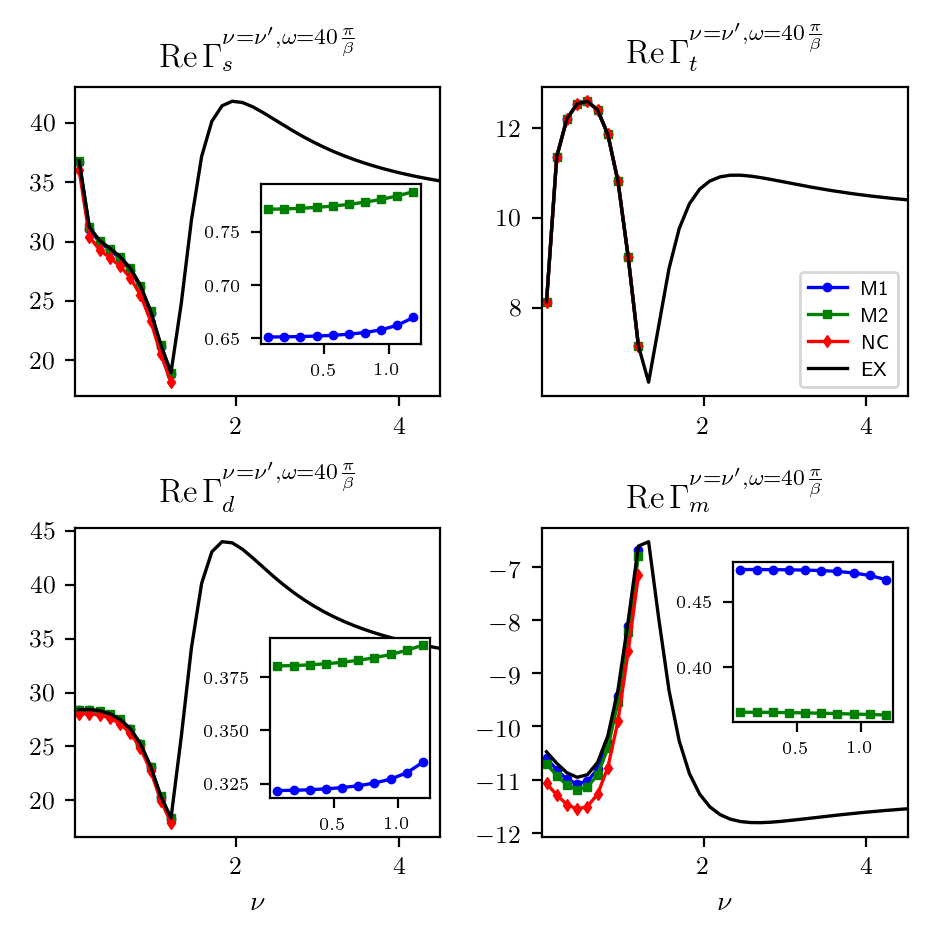}}
		\caption{[Color online]  $\Gamma_r^{\nu\nu'\omega}$ for $r=\lbrace s,t,d,m \rbrace$, evaluated along the diagonal $\nu = \nu'$ for $U=1.75$, $\omega=0$ (left panels) and $\omega=40\pi/\beta$ (right panels). Results of method $1$ (blue) and method $2$ (green) are compared to a plain inversion (red) of Eq.~(\ref{equ:calcGamma}) for $N_{\text{inv}}\!=\!40$ and the exact solution (black). Insets show the corrections terms only. }
\label{fig:comparison_gamma_U1p75}
\end{center}
\end{figure}

In this Appendix, we report the behavior of $\Gamma^{\nu \nu' \omega}$ for $U=1.75$ (see Fig. \ref{fig:comparison_gamma_U1p75}). We compare the data obtained by the two methods to correct the inversion of the Bethe-Salpeter equations with to the ``non-corrected" result. Differently from the case $U=1$, one observes, for the singlet and the density channels, a non-negligible fermionic frequency structure of the corrections provided by the two methods. Albeit quite relevant for the case $N_{\text{inv}}\!=\!40$, this structure has been shown to became negligible going to higher values of $N_{\text{inv}}$. By looking at the fermionic frequency dependence provided by the second line of Eq.~(\ref{equ:gamma00final1}) (method $1$) and Eq.~(\ref{equ:gamma00final2}) (method $2$), one can deduce (see discussion in Sec.~\ref{sec:num_results}) that this is originated by the asymptotics of $\Gamma_{r, \text{asym}}$ in Eq.~(\ref{equ:Gammahighfreq}). In the intermediate regime, one would expect the low-frequency structure of $\Gamma_r$ (not captured by the asymptotics) to become particularly pronounced in the density and the singlet channels \cite{Schaefer2013,Chalupa2018}. Possibly, this creates a cancellation with the (fermionic) frequency-dependent terms of $\Gamma_{r, \text{asym}}$ in the regime where the low-frequency structure in not fully decayed. As one can see in Fig.~\ref{fig:vertex} the latter are not fully decayed for $N_{\text{inv}}\!=\!40$. This can explain the (fermionic)frequency structure shown in the insets if Fig.~\ref{fig:comparison_gamma_U1p75}, which can be reduced by considering a bigger $N_{\text{inv}}$ range.

\twocolumngrid

\end{document}